\documentclass[slac_one]{revtex4}
\usepackage{graphicx}
\usepackage{fancyhdr}
\def\MSbar{\hbox{\tiny ${\overline{\rm MS}}$}}
\def\eq#1{Eq.~(\ref{#1})}
\catcode`\@=11 
\def\lsim{\mathrel{\mathpalette\@versim<}}
\def\gsim{\mathrel{\mathpalette\@versim>}}
\def\@versim#1#2{\vcenter{\offinterlineskip
        \ialign{$\m@th#1\hfil##\hfil$\crcr#2\crcr\sim\crcr } }}
\catcode`\@=12 

\begin{document}

\title{Inclusive distributions near kinematic thresholds}


\author{Einan Gardi}

\affiliation{Cavendish Laboratory, University of Cambridge, J J
Thomson Avenue, Cambridge, CB3 0HE, UK\\}
\affiliation{Department of Applied Mathematics \& Theoretical Physics,
Wilberforce Road, Cambridge CB3 0WA,~UK}
\begin{abstract}
The main challenge in computing inclusive cross sections and decay
spectra in QCD is posed by kinematic thresholds.
The threshold region is characterized by stringent phase--space constraints
that are reflected in large perturbative corrections due to
soft and collinear radiation as well as large non-perturbative effects.
Major progress in addressing this problem was made in recent years by
Dressed Gluon Exponentiation (DGE), a formalism that combines Sudakov
and renormalon resummation in moment space.
DGE has proven effective in extending the range of applicability of
perturbation theory well into the threshold region and in identifying the
relevant non-perturbative corrections.
Here we review the method from a general perspective using examples
from deep inelastic structure functions, event--shape distributions,
heavy--quark fragmentation and inclusive decay spectra.
A special discussion is devoted to the applications of DGE to radiative
and semileptonic B decays that have proven valuable for the interpretation
of data from the B factories.
\end{abstract}

\maketitle

\thispagestyle{fancy}

\section{Introduction}

The calculation of inclusive differential cross sections and decay
spectra is amongst the most important and well--developed
applications of perturbative QCD. The main challenge in computing
such distributions arises from kinematic thresholds, and specifically from
the exclusive boundary of phase space.
The threshold region is characterized by stringent constrains
on real gluon emission,
leading to large perturbative and non-perturbative corrections.
This region is important for phenomenology
in a wide range of applications.
Well known examples are deep inelastic structure
functions in the limit where
Bjorken $x$ approaches~1~
\cite{Sterman:1986aj}--\cite{GR};
event--shape distributions in $e^+e^-$ annihilation
near the two--jet
limit~
\cite{Sterman:1977wj}--\cite{BM};
Drell--Yan~\cite{CMW,Sterman:1986aj,CT}, \cite{Gardi:2001di}--\cite{Laenen:2000ij}
or Higgs~\cite{Kulesza:2003wn,Bozzi:2005wk} production in hadronic
collisions near partonic threshold, or at small transverse momentum;
heavy--quark fragmentation~\cite{Nason:1996pk}--\cite{Gardi:2003ar}
and inclusive decay
spectra~\cite{Neubert:1993um}--\cite{Gardi:2006gt}.
In all these examples resummation and
identification of the relevant non-perturbative corrections can
significantly increase the range of applicability of perturbation
theory.

Stringent constraints on real gluon emission are reflected in
large Sudakov logarithms~\cite{Sudakov:1954sw}, namely perturbative
corrections associated with mass singularities that become
parametrically large near the exclusive limit. Owing to
factorization, Sudakov logarithms can be resummed. It is generally
the case, however, that upon approaching the threshold
(while keeping the hard momentum scale $Q$ fixed) non-perturbative effects
become dominant. Sudakov resummation is therefore useful in a
restricted range {\em that is bounded from both ends}: the
logarithms need to be large enough to dominate, but in the close
vicinity of the threshold, where the logarithms are indeed
very large, the perturbative expansion as a whole breaks down, and
it no longer approximates the physical distribution.

At the perturbative level, a typical infrared and collinear
safe~\cite{Sterman:1977wj}
differential cross section\footnote{Although the problem addressed here
is completely general, for simplicity we consider an
infrared and collinear safe distribution, where collinear factorization is
not needed. We further simplify the discussion assuming a single differential
distribution where the hard kinematics is fixed, such as the thrust ($T$)
distribution
in $e^+e^-$ annihilation~\cite{Gardi:2002bg} where $Q$ is the
center--of--mass energy and
$x\equiv T$ (the threshold, $x=1$, corresponds to the two--jet limit)
or the photon--energy spectrum in
$\bar{B}\longrightarrow X_s\gamma$~\cite{Andersen:2005bj}
where $Q=m_b$ and $x\equiv 2E_{\gamma}/m_b$ (the threshold,
$x=1$, corresponds to the maximal value of $E_{\gamma}$ for an
on-shell b quark decay).}
(or decay spectrum) takes the form
\begin{equation}
\label{pert_mom} \frac{1}{\sigma_{\rm tot}}\frac{d\sigma
(Q,x)}{dx}= H(\alpha_s(Q^2)) \delta(1-x) \,+\,R(\alpha_s(Q^2),x),
\end{equation}
where $x$ is some measured momentum fraction, the distribution has
support for $x\leq 1$ and $x=1$ is the threshold where the
leading order distribution $\delta(1-x)$ receives purely virtual
corrections, $H(\alpha_s)=1 +v_1\alpha_s+\ldots$.
\hbox{$R(\alpha_s,x)=r_1(x)\alpha_s+\ldots$} contains terms of the
form $\left[\ln^k(1-x)/(1-x)\right]_+$, where the plus distribution
is defined by
\begin{equation}
\label{plus} \int_0^1 dx f(x) \left[\frac{\ln^k(1-x)}{1-x}\right]_+
=\int_0^1 dx \Big(f(x)-f(1)\Big) \frac{\ln^k(1-x)}{1-x},
\end{equation}
where $f(x)$ is a smooth test function. These distributions
incorporate the cancellation between real and virtual mass
singularities that are associated with the limit where a parton is
fully resolved. At order $\alpha_s^n$ Sudakov logarithms appear with
$0\leq k\leq 2n-1$. These perturbative terms get large as
$x\longrightarrow 1$ even if the coupling is small, and therefore
resummation is necessary.

The problem of the threshold region is best
formulated in moment space,
\begin{equation}
\label{mom} \tilde{\sigma}(Q,N)=\int_0^1\, dx \,x^{N-1} \,
\frac{1}{\sigma_{\rm tot}}\frac{d\sigma (Q,x)}{dx},
\end{equation}
where the distributions become analytic functions of the moment
index $N$. High spectral moments
 in \eq{mom} are increasingly sensitive
to the limit $x\longrightarrow 1$; the plus distributions
described above give rise to logarithmically--enhanced terms,
$\ln^k (N)$, which dominate at $N\longrightarrow \infty$.
Moment space is useful
for Sudakov resummation because the multi--parton phase space
factorizes there up to ${\cal O}(1/N)$ corrections.
Factorization properties of QCD matrix elements
for soft and collinear radiation can then be used to prove
\emph{exponentiation}~\cite{Sudakov:1954sw}--\cite{Contopanagos:1996nh}: all the singular terms,
$\alpha_s^n\, \ln^k (N)$, are generated by
\begin{equation}
\label{mom_sud} \tilde{\sigma}(Q,N)=H(\alpha_s(Q^2))\,\times \,
{\rm Sud}(Q,N)\,+\,{\cal O}(1/N),
\end{equation}
where the Sudakov factor takes the form
\begin{equation}
\label{Sud_expansion}
{\rm Sud}(Q,N)=\exp\bigg\{\sum_{n=1}^{\infty}
\sum_{l=1}^{n+1} E_{n,l} \,\alpha_s^n(Q) \,\ln^l(N)
\bigg\},
\end{equation}
where $E_{n,l}$ are numerical coefficients, which depend on the relevant
Sudakov anomalous
dimensions and the coefficients of the $\beta$ function.
The resummed moment--space expression of \eq{mom_sud}
can readily be matched to the fixed--order result to determine
$H(\alpha_s(Q^2))$ and to account for the
${\cal O}(1/N)$ contributions. Finally, the resummed cross section is
obtained by an inverse Mellin transformation:
\begin{equation}
\label{inv_Mellin}
\left.
\frac{1}{\sigma_{\rm tot}}\frac{d\sigma (Q,x)}{dx}\right\vert_{\rm res}
= \int_{c-i\infty}^{c+i\infty}\frac{dN}{2\pi\,i}\,
\tilde{\sigma}(Q,N),
\end{equation}
where the integration contour passes to the right of all the singularities
of $\tilde{\sigma}(Q,N)$.

The physics underlying this resummation
is the quantum--mechanical incoherence of dynamics
at different momentum scales. This general principle is not restricted
to the perturbative level
--- an observation that will be important for what follows.
The typical excitations that are relevant at large $N$ have virtualities
of ${\cal O}(Q^2/N)$ ---
the jet--mass scale or ${\cal O}(Q/N)$ --- the ``soft scale''.
Because soft or collinear gluons do not resolve the hard interaction, they
decouple from the process--dependent dynamics at the hard
scale $Q$.

\begin{figure}[t]
\includegraphics[width=125mm]{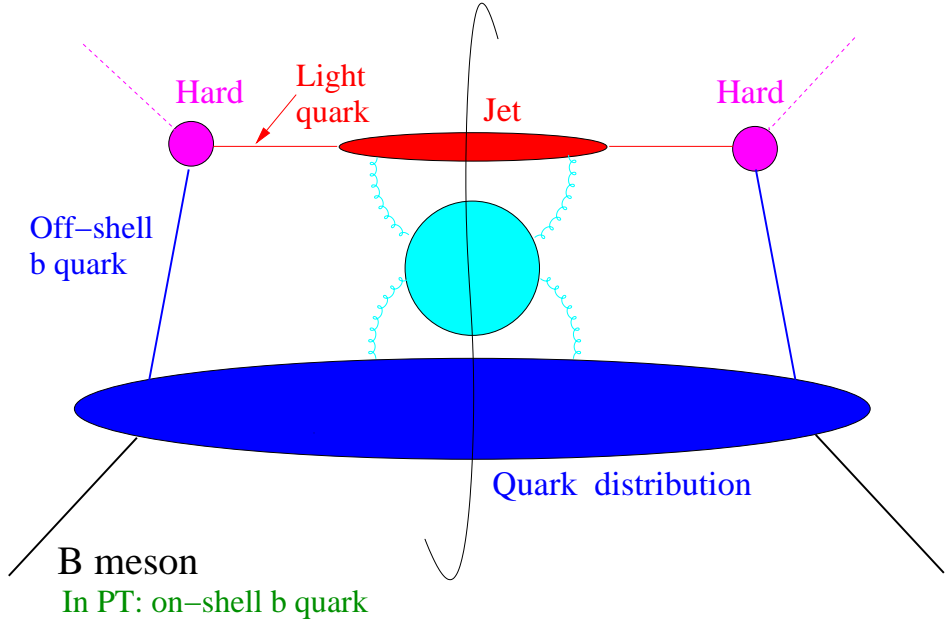}
\caption{Factorization of inclusive decays into ``hard'', ``jet'' and
 ``soft'' functions, with virtualities of order $m_b$,
$m_b/\sqrt{N}$ and $m_b/N$, respectively.
The jet function describes an unresolved jet with a given mass;
see Ref.~\cite{GR} and Sec.~\ref{DIS} below.
In inclusive decays
 the ``soft'' function $S_N(m_b;\mu_F)$ has the process--independent interpretation of a quark
 distribution function. This function describes the longitudinal momentum distribution
of an off-shell b quark field in the initial state (an on-shell b quark in perturbation
theory, and a ${\bar B}$ meson in the full theory).
See Ref.~\cite{QD} for a precise definition.
 \label{fact_B_decay}}
\end{figure}

For $N\longrightarrow \infty$ the moments $\tilde{\sigma}(Q,N)$
factorize into ``hard'', ``jet'' and ``soft''
functions~\cite{Sterman:1986aj,CT,Collins:1984kg,Laenen:2000ij,Korchemsky:1994jb}\footnote{In the
context of inclusive decays,
see also Refs.~\cite{Bauer:2000yr,Bosch:2004th,Andersen:2005bj}.}.
Upon introducing a factorization scale $\mu_F$,
the Sudakov factor in \eq{mom_sud} can be written as
\begin{equation}
{\rm Sud}(Q,N)=J_N(
Q;\mu_F)\,\times\,S_N(Q;\mu_F),
\label{factorization}
\end{equation}
where the jet function
$J_N(Q;\mu_F)$ and the
``soft'' function $S_N(Q;\mu_F)$ sum up Sudakov logarithms associated with
the jet--mass scale $Q^2/N$ and with the ``soft scale''
$Q/N$, respectively. The factorization--scale dependence\footnote{Factorization
is implemented here using dimensional regularization
(the $\overline {\rm MS}$ scheme). In contrast
with~\cite{Sterman:1986aj,Laenen:2000ij},
we do not employ strict scale separation:
both $J_N(Q;\mu_F)$ and  $S_N(Q;\mu_F)$ have some residual dependence
on the hard scale $Q$; these functions are normalized such that their first
moment is unity: $J_{N=1}(Q;\mu_F)\equiv 1$ and $S_{N=1}(Q;\mu_F)\equiv 1$.
}
cancels
exactly in the product in \eq{factorization}.
The phase--space origin of the logarithms
is illustrated in Fig.~\ref{fact_B_decay} in the example
of inclusive B decays.

Sudakov logarithms, reflecting the hierarchy
$Q/N\, \ll\, Q/\sqrt{N}\, \ll \, Q$, can be resummed to all orders.
One expects, however, that a purely perturbative treatment
will only be valid for $Q/N\gg
\Lambda$. Even then, extremely soft gluons with virtualities
 ${\cal O}(\Lambda)$, whose interaction is not described
by perturbation theory,
have some effect on the distribution. These gluons become increasingly important
as the hierarchy $Q/N\gg \Lambda$ is removed.
Considering the large--$N$ limit for fixed $Q$,
one finds that the threshold region is characterized not only by
Sudakov logarithms but also by
non-perturbative effects that are inversely proportional to
the ``\emph{soft scale}'',
powers of $N\Lambda/Q$, which eventually \emph{always get
larger than the logarithms as $N$ increases}.

Importantly, the perturbative
calculation itself reflects the presence of parametrically--enhanced
power corrections:
the series in the exponent of \eq{Sud_expansion} is non-summable owing to
infrared renormalons~\cite{DGE_thrust,Gardi:2001di}.
In conventional Sudakov resummation this infrared sensitivity is ignored:
the perturbative sum is truncated, not regularized.
A regularization is necessary in order to systematically
separate between logarithmically--enhanced perturbative corrections
and power--enhanced non-perturbative power corrections.
Such power corrections are uniquely associated with the phase--space region
from which the logarithms arise, and therefore their physical interpretation
is straightforward. For example, in event--shape distribution the all--order
resummation of (large--angle) soft gluon emission exposes
parametrically--enhanced \emph{hadronization corrections};
in inclusive decay spectra (Fig.~\ref{fact_B_decay})
all--order soft--gluon resummation exposes
parametrically--enhanced corrections distinguishing
the quark distribution in the initial--state meson from that in an
on-shell heavy quark.

The presence of infrared renormalons and parametrically--enhanced
power corrections
in the moment--space Sudakov factor is completely general, and it has
far--reaching implications for the analysis of inclusive distributions.
On general grounds, independently of the renormalon perspective,
treatment of power corrections on the
soft scale for $N\sim Q/\Lambda$ requires the
introduction of \emph{a new non-perturbative
function}~\cite{Neubert:1993um,Neubert:1993ch,Bigi:1993ex,Korchemsky:1999kt},
the so-called ``shape function''.
Unfortunately, the ``shape function'' is hard to constrain theoretically, while
if constrained by data alone the result would strongly depend on the
functional form assumed and on the perturbative approximation with
which it is combined.

In practice, the threshold region was addressed by different tools
in different applications, usually introducing some infrared cutoff on the
perturbative result and parametrizing the contributions from the infrared
as a ``shape function''. Having little theoretical constraints on this
function, the predictive power has been, in general, very limited.
The most familiar examples are:
\begin{itemize}
\item{} {\bf Event--shape distributions:} Hadronization effects are
taken into account though a
leading non-perturbative ``shift'' of the Sudakov resummed
spectrum~\cite{Dasgupta:2003iq,Dokshitzer:1997ew,Korchemsky:1995zm}
or, when addressing the peak region, through a
convolution with a ``shape function''
\cite{Korchemsky:1999kt,Korchemsky:2000kp}. The general strategy of
quantifying hadronization effects by power corrections represents an
important advancement with respect to their description through
fragmentation models in event generators. This approach quickly led
to successful phenomenology for average values of event
shapes~\cite{Korchemsky:1995zm}--\cite{Milan},~\cite{Gardi:1999dq},
but the challenge of describing the spectrum in the peak region
could not easily be met.
\item{} {\bf Heavy--meson production cross sections:}
 Fragmentation of the heavy quark into a meson
is taken into account through a
 convolution with a leading--power fragmentation
function~\cite{Kart,Peterson}. Applied with or without Sudakov
resummation, usually without any momentum cutoff, see
e.g.~\cite{Cacciari:2005uk}.
\item{} {\bf Inclusive B decay:}
The momentum distribution of the b quark in the B meson is taken
into account through a convolution with a leading--power ``shape
function''~\cite{Neubert:1993um,Neubert:1993ch,Bigi:1993ex,Bosch:2004th,Lange:2005yw}. Applied
with or without Sudakov resummation, using a momentum cutoff.
\end{itemize}

Major progress in addressing this problem from first principals was
made in recent years by Dressed Gluon Exponentiation (DGE). In DGE
the {\em moment--space} Sudakov exponent is computed as an {\em
all--order Borel sum}, avoiding the usual logarithmic--accuracy
truncation. {\em Power--like separation} between perturbative and
non-perturbative corrections on the ``soft scale'' $Q/N$ is achieved
by the Principal Value prescription, avoiding any arbitrary momentum
cutoff. This opens the way for making full use of the inherent
infrared safety of the observable, which often extends beyond the
logarithmic level. Infrared factorization and renormalization--group
invariance strongly constrain the universal, all--order structure of
the exponent. Additional, observable--dependent properties are
derived from renormalon calculations of the corresponding Sudakov
anomalous dimensions. These translate into {\em constraints on the
parametric form} of power corrections that affect the exponentiation
kernel. The end result is that the physical spectrum can be
approximated by the regularized perturbative DGE spectrum,
supplemented by {\em a few} power corrections of a known parametric
form. This highly predictive framework has led to successful
phenomenology in a variety of applications, including in particular
event--shape distributions~\cite{DGE_thrust,Gardi:2002bg},
heavy--quark fragmentation~\cite{CG} and inclusive B decay
spectra~\cite{Andersen:2005bj,Andersen:2005mj}.

Our purpose here is to review the DGE approach from a
general perspective. We will discuss the
applications mainly to illustrate how the method works
and distinguish between general and process--specific
properties.
We start, however, in Sec. \ref{BDK} by shortly describing the
motivation and the present status of the calculation of inclusive B
decay spectra, where the application of DGE has proven useful
for the interpretation of data from the B factories
~\cite{BDK,Andersen:2005bj,Gardi:2005mf,Andersen:2005mj,Gardi:2006gt}. In
Sec. \ref{DIS} we explain the foundation of the DGE approach using the
example of an unresolved jet. This analysis~\cite{GR} is directly relevant for the
 problem of deep inelastic structure functions at large Bjorken $x$, where
the power corrections can also be viewed as resummation of
the Operator Product Expansion (OPE)~\cite{Gardi:2002bk}.
Next, in Sec.~\ref{SOFT}, we turn to discuss large--angle soft radiation.
We explain how constraints on power corrections
follow from the properties of the corresponding Sudakov anomalous dimension.
The success of the resulting power--correction phenomenology is demonstrated
using examples from event--shape
distributions, heavy--quark fragmentation and B decay spectra.
This is followed by a short summary of our conclusions in Sec.~\ref{Conc}.

\section{Inclusive B decay spectra\label{BDK}}

Amongst the most important contributions of the B factories are
inclusive B decay measurements~\cite{Barberio:2006bi,HFAG}:
\begin{itemize}
\item{} inclusive semileptonic decays, which
provide the most accurate way to determine the CKM parameters
$|V_{\rm cb}|$ and $|V_{\rm ub}|$. The ratio $|V_{\rm ub}/V_{\rm cb}|$ is one of
the important constraints on the unitarity
triangle~\cite{Bona:2005vz,Charles:2004jd}. It is also
an essential input in probing new physics, for example through
$B^0$--$\bar B^0$ mixing,
see e.g.~\cite{Ball:2006xx}.
\item{} inclusive radiative decay, ${\bar B}\longrightarrow X_s \gamma$,
that directly probes the short--distance interaction responsible for
flavor changing neutral currents.
This decay, which occurs only through loops within the Standard Model,
provides an important constraint on possible new physics
scenarios~\cite{Allanach:2006jc,Gambino:2005dp}.
\end{itemize}
Inclusive decays are theoretically favorable over exclusive ones
owing to their low sensitivity to the hadronic structure of the initial
and final states.

Owing to irreducible backgrounds, experimental measurements of
inclusive decays are limited to certain kinematic regions.
This makes the theoretical calculation of inclusive decay
\emph{spectra} an essential ingredient on the way from the
measurement to its interpretation. This is the case for both the
radiative and the charmless semileptonic decays; the exception is
${\bar B}\longrightarrow X_c l \bar{\nu}$ that
can be measured over the whole kinematic range.
In ${\bar B}\longrightarrow X_s \gamma$, only
hard photons with $E_{\gamma}>1.8$ GeV, corresponding to
$M_X\equiv \sqrt{(P_B-q)^2}=\sqrt{M_B(M_B-2E_{\gamma})}\lsim 3$ GeV,
can be distinguished from the background; the measurement
becomes statistically limited only for $E_{\gamma}\gsim 2.1$ GeV.
For the charmless semileptonic decay,
${\bar B}\longrightarrow X_u l \bar{\nu}$,
the situation is yet more difficult:
an upper cut on the hadronic invariant mass $M_X<1.7$ GeV must be
applied to avoid the overwhelming charm background.
Thus, the determination of $|V_{\rm ub}|$ from inclusive measurements strongly
relies on the theoretical calculation of the spectrum.

\begin{figure}[t]
\includegraphics[width=95mm]{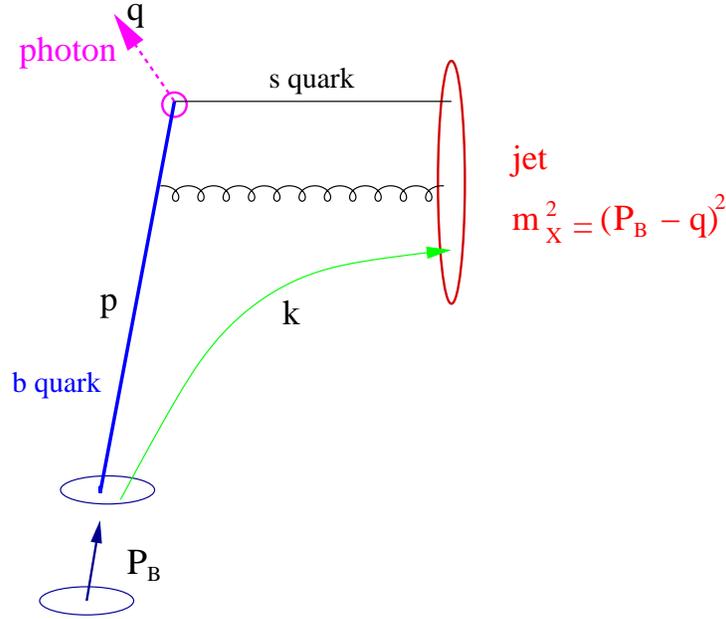}
\caption{The kinematics in ${\bar B}\longrightarrow X_s \gamma$:
the photon, carrying momentum $q$ (with $q^2=0$), recoils against
the hadronic jet.
In general, the virtuality of the b quark when decaying
is a consequence of its ``primordial''
Fermi motion as well as of its subsequent interaction.
In the on-shell approximation, where perturbation theory applies,
the ``primordial''
Fermi motion, namely the interaction of the b quark with the
light degrees of freedom in the meson is neglected;
then the $b$ quark momentum in the meson rest frame is
$p=(m_b,\vec{0})$ while the momentum of the light degrees of freedom is
$k=(\bar{\Lambda},\vec{0})$, with
$\bar{\Lambda}=M_B-m_b$, where $M_B$ is the meson mass and $m_b$ is
the quark pole mass.
 \label{kin}}
\end{figure}

As dictated by the Born--level process, the typical hadronic momentum
configuration in heavy--to--light decays is \emph{jet--like}
(see Fig.~\ref{kin}):
in the B rest frame the hadronic system has a high energy and a small mass,
$M_X\ll M_B$. The spectra peak near threshold\footnote{For example,
the ${\bar B}\longrightarrow X_s \gamma$ spectrum peaks near
$E_{\gamma}\simeq m_b/2$, see Fig.~\ref{DGE_vs_FO} below.},
making the understanding of this
limit absolutely essential. As discussed above,
only the small--$M_X$ region is experimentally accessible.

An important observation underlying the theoretical description of
inclusive heavy--to--light decay spectra, is that the decaying
b quark is not on-shell~\cite{Neubert:1993um,Neubert:1993ch,Bigi:1993ex}.
The Fermi motion of the b quark in the meson involves momenta of
${\cal O}(\Lambda)$, and therefore one can expect ${\cal O}(\Lambda)$
smearing of the perturbative spectrum by non-perturbative effects.
The common lore, which developed based on this physical picture, is that
computing the spectrum in the peak region is strictly beyond the limits of
perturbative QCD.

\begin{figure}[t]
\includegraphics[width=85mm]{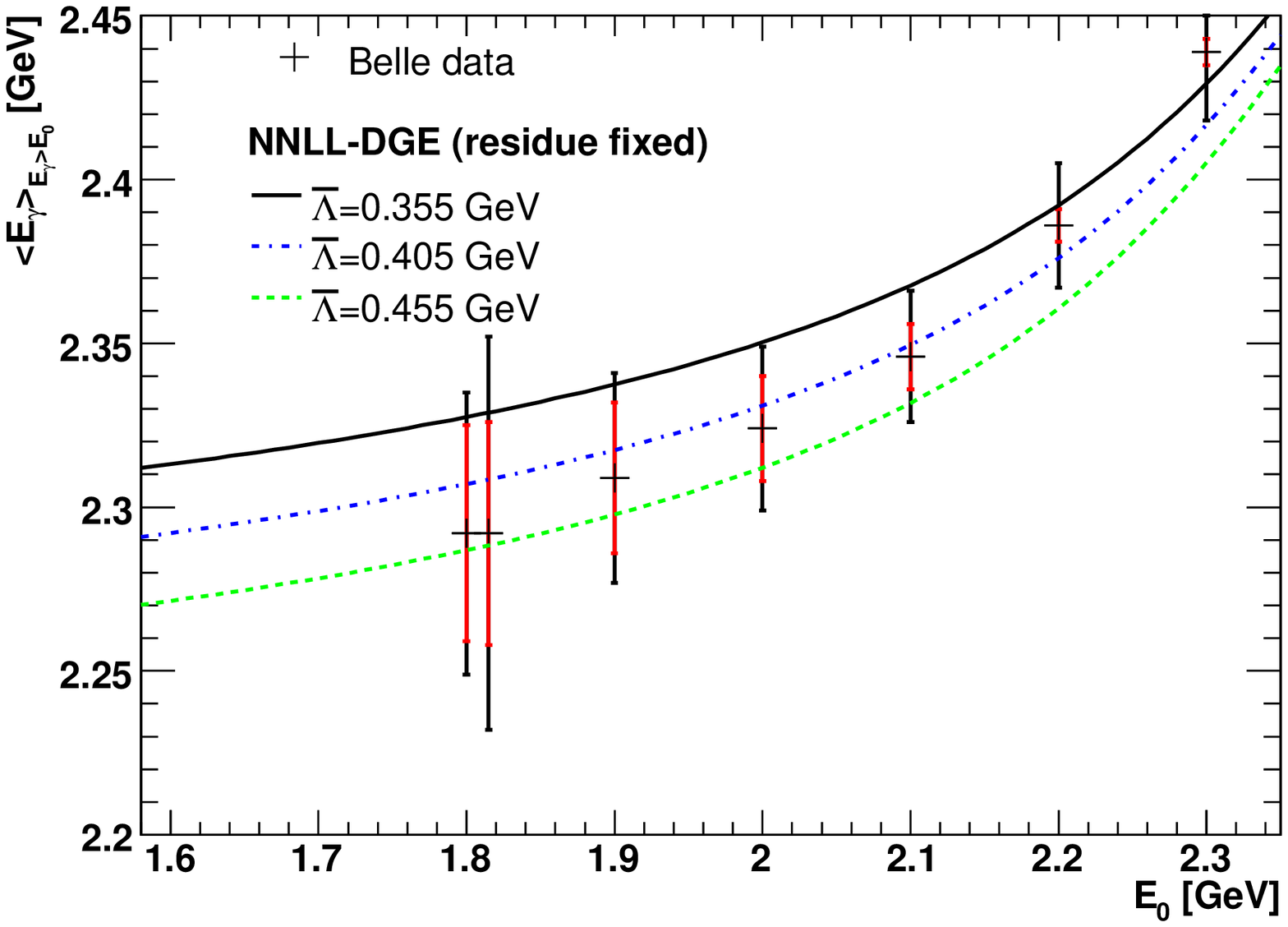}
\includegraphics[width=85mm]{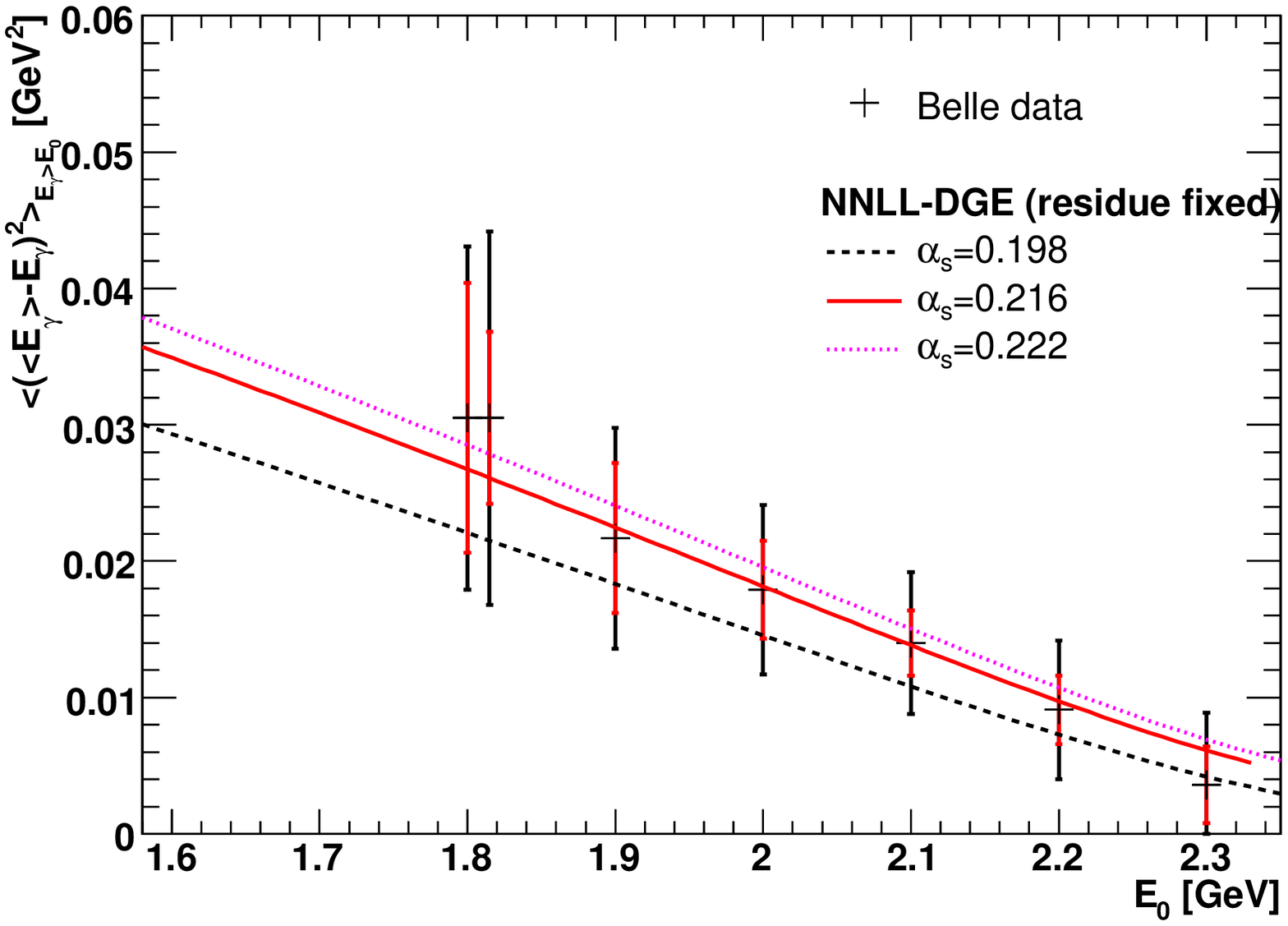}
\caption{$\bar{B}\longrightarrow X_s\gamma$ theory {\it{vs.}} data:
The average energy (left) and the variance (right) with a cut $E_{\gamma}>E_0$,
as calculated by DGE in~\cite{Andersen:2005bj} in the on-shell approximation
(i.e. with no power corrections),
varying the short distance parameters $m_b^{\MSbar}$ and $\alpha_s^{\MSbar}$ within
their error range. The result is
compared with data from Belle~\cite{Belle05}. Inner and total error
bars show systematic and statistical plus systematic errors (added
in quadrature), respectively.
 \label{belle_moments}}
\end{figure}
In fact, the situation is significantly
better~\cite{Andersen:2005bj,Gardi:2005mf,Andersen:2005mj,Gardi:2006gt}:
a systematic on-shell calculation
involving resummation of the perturbative expansion yields a good approximation
to the meson decay spectrum, one which provides an excellent
starting point for quantifying non-perturbative corrections.

The on-shell approximation is
physically natural because the heavy
quark carries most of the momentum of the meson: the b quark
virtuality, ${\cal O}(\Lambda)$, is much smaller than the mass.
When considering the total rate this translates into a systematic
expansion in inverse powers of the mass, where the leading corrections
are ${\cal O}(\Lambda^2/m_b^2)$ and are numerically small.

The next, crucial observation
is that the on-shell decay spectrum is \emph{infrared and collinear safe},
namely
{\em
all its moments have finite coefficients to any order in perturbation theory}.
This means that non-perturbative effects, which make for the difference
between the on-shell approximation and the physical meson decay,
appear in moment space only through {\em power corrections.}
Of course, when referring to the on-shell decay spectrum one must address
the question of the summability of the expansion (as well as the precise definition
of the on-shell mass), which brings about the issue of infrared
renormalons.

In general, the virtuality of the decaying b quark
is a consequence of its ``primordial''
Fermi motion and of its subsequent interaction.
This, however, does not
preclude the resummed on-shell spectrum being a good approximation to the
meson decay spectrum. Obviously, the ``primordial'' Fermi motion and the
``subsequent interaction'' need to be defined. This definition amounts
to a separation between non-perturbative and perturbative contributions, which
can be done in a variety of ways.
Cutoff--based separation
procedures~\cite{Neubert:1993um}--\cite{Lange:2005yw}, \cite{Bigi:1996si}
that are often applied, hinder the possibility of using of the inherent
infrared safety of the on-shell decay spectrum. They necessarily associate
a significant contribution with the non-perturbative ``primordial''
Fermi motion, which must be parametrized in a so-called ``shape function''.
Having little theoretical guidance, this parametrization is rather arbitrary,
and the predictive power is very limited.
The alternative used in DGE is separation at the level of powers, which is
implemented using Principal Value Borel summation. In this way one can make full
use of the inherent infrared safety of the on-shell decay spectrum, leaving only
genuinely non-perturbative contributions, which distinguish between the
(properly regularized) quark distribution in an on-shell quark and that in
a meson, to be parametrized. Moreover, this parametrization is well guided by the
theory: a small number of non-perturbative power corrections of a
known form (see below) would suffice. In fact, with present experimental data
no power corrections are needed --- see Fig.~\ref{belle_moments}.

It must be emphasized that the on-shell decay spectrum can
only be an approximation to the physical spectrum when considered {\em to
all orders}. In particular, both
the on-shell mass, which sets the perturbative endpoint at any
order in perturbation theory
$E_\gamma\longrightarrow m_b/2$, or
$x\equiv 2E_{\gamma}/m_b\longrightarrow 1$, and the spectral moments
defined with respect to $x$ as in \eq{mom},
have a leading infrared renormalon
at $u=\frac12$. The corresponding
ambiguity cancels upon
computing the sum of the series
for $m_b$ and for the Sudakov factor using
a systematic regularization --- see \eq{cancellation} below.
Examining the real--emission contribution to the spectrum in fixed--order
perturbation theory (Fig.~\ref{DGE_vs_FO}) one indeed observes huge
corrections\footnote{The `no resummation' curves in
Fig.~\ref{DGE_vs_FO} should not be considered too seriously for
several reasons. First, the pole mass used at each order is the same, while
in a strictly fixed--order treatment, it would be different
(it diverges quickly).
Second, the coupling is arbitrarily renormalized at $m_b$,
while the typical gluon virtuality
(the BLM scale~\cite{BLM,Brodsky:2000cr}) is significantly lower.} going from
order to order\footnote{A full NNLO calculation of the normalized spectrum corresponding to the
electromagnetic dipole operator
was performed in Ref.~\cite{MM}.}. Considering the resummed spectrum by DGE,
matched to
NLO and NNLO, respectively~\cite{Andersen:2005bj,Andersen_Gardi_to_appear}, one
 finds instead a remarkable stability,
reflecting the fact that the dominant
corrections have been resummed. Another striking difference between
the resummed spectrum and the fixed--order one is their different support
properties: the fixed--order result is a \emph{distribution} that has
support for $E_{\gamma}\leq m_b/2$, while the resummed result is a
 \emph{function}
that has support for
 {\hbox{$E_{\gamma}\leq (m_b+{\cal O}(\Lambda))/2\simeq M_B/2$}}, i.e.
close to the physical spectrum. Evidently, resummation makes a qualitative
difference.
\begin{figure}[t]
\includegraphics[width=85mm,angle=90]{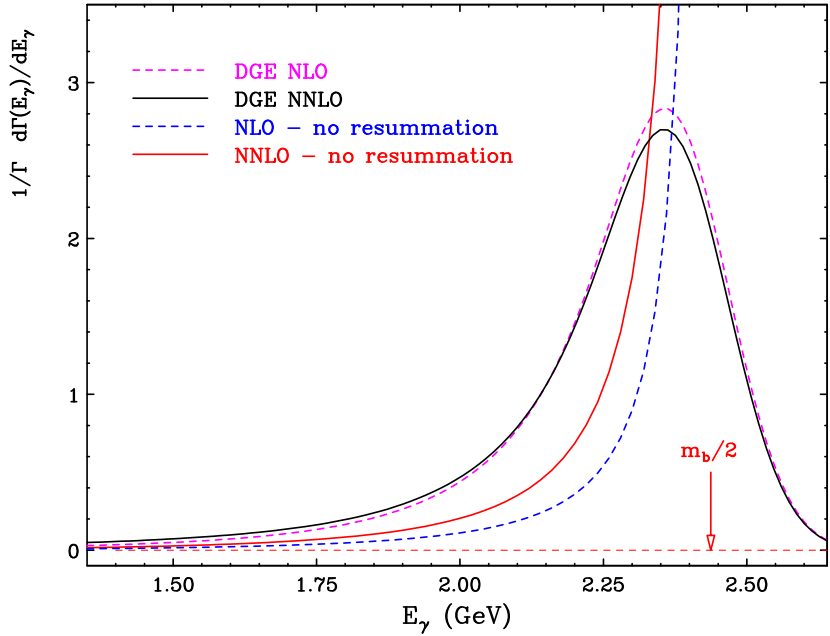}
\caption{The $\bar{B}\longrightarrow X_s\gamma$ photon--energy spectrum
(corresponding to the electromagnetic dipole operator $O_7$)
in the on-shell
approximation as computed in fixed--order perturbation theory at NLO and at
NNLO,
and by resummed perturbation
theory (DGE)~\cite{Andersen:2005bj,Andersen_Gardi_to_appear}, matched to
NLO and NNLO, respectively. A common value of the (Principal Value) pole mass,
$m_b=4.88\pm 0.05$ GeV, has been used. This value was
computed~\cite{Andersen:2005bj} based
on the measured short--distance quark mass in~$\overline{\rm MS}$,
$m_b^{\MSbar}=4.20\pm 0.04$~GeV \cite{Buchmuller:2005zv}.
 \label{DGE_vs_FO}}
\end{figure}

The ultimate test of the theoretical predictions is of course the
comparison with experimental data. In $\bar{B}\longrightarrow X_s \gamma$ spectral
data from Babar and Belle has become available shortly after the publication
of the theocratical predictions by DGE~\cite{Andersen:2005bj}.
Comparison of the first two spectral
moments, computed with a varying lower cut $E_{\gamma}>E_0$,
with Belle data is shown in Fig.~\ref{belle_moments};
a similar comparison with BaBar data can be found in Ref.~\cite{Gardi:2005mf}.
These results show that the resummed on-shell decay spectrum
indeed provides a good
approximation to the meson decay spectrum, as anticipated in
Ref.~\cite{Andersen:2005bj}.
Comparison of the computed $\bar{B}\longrightarrow X_s \gamma$ spectrum
with data has several immediate applications:
\begin{itemize}
\item{} Extrapolation of the
partial width from the region of measurement to the whole of phase space,
in order to confront it with the Standard Model prediction and provide
constrains on physics beyond the Standard Model.
\item{} Precise determination of the b quark mass, see Ref.~\cite{HFAG}.
\item{} Determination of the leading power corrections,
the ones associated with the quark distribution in the meson.
With the present accuracy, power corrections
cannot be determined, but higher precision data are expected in
the near future. Owing to their universality, these power corrections can
readily be used in improving theoretical predictions for other inclusive
decay spectra, notably the charmless semileptonic spectrum.
\end{itemize}

\begin{figure}[t]
\includegraphics[width=125mm,angle=0]{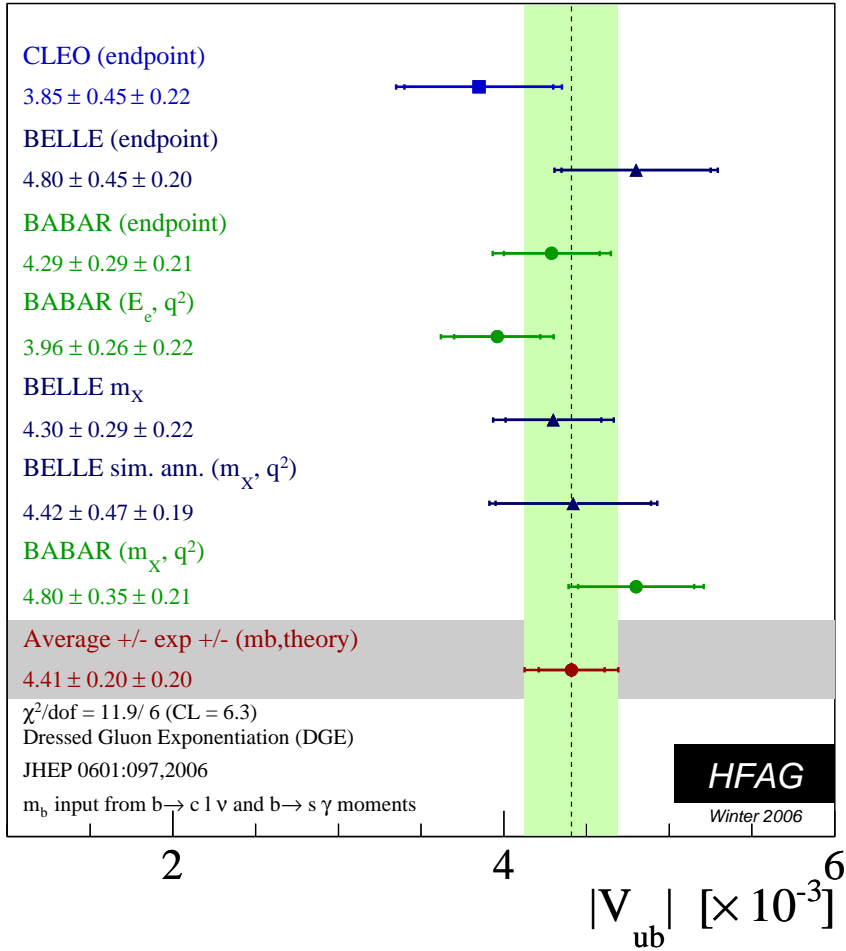}
\caption{$|V_{\rm ub}|$ as extracted using DGE~\cite{Andersen:2005mj}
by the Heavy Flavor Averaging Group (HFAG)~\cite{HFAG}
from inclusive semileptonic
measurements by CLEO, Belle and BaBar.
\label{HFAG_Vub}}
\end{figure}
A pressing issue, which is high on the agenda of the B factories is the precise
determination of $|V_{\rm ub}|$. As mentioned above, inclusive measurements
of charmless semileptonic decays
provide
the most promising avenue for this determination. The main obstacle
is the large extrapolation from the
region of measurement to the whole of phase space,
which strongly relies on theoretical predictions for the spectrum.
The strategy applied so far~\cite{Barberio:2006bi} made direct use of
the universality of the quark distribution in the meson,
by first fitting an ansatz for the ``shape function'' to the
$\bar{B}\longrightarrow X_s \gamma$ decay data and then using it when
computing the partial branching fraction of
$\bar{B}\longrightarrow X_u l \bar{\nu}$ within the region of measurement.
As discussed above, in this approach
there is little theoretical guidance on the functional form
of the ``shape function''.
Here the alternative presented by DGE is very
attractive~\cite{Andersen:2005mj,Gardi:2006gt}:
in this framework the on-shell calculation, which depends only on the
short--distance parameters, provides a good
 approximation to the spectrum. Moreover, prospects are high for improving this prediction
further by higher--order (NNLO) calculations and by parametrization of
the first few power corrections whose $N$--dependence is known.

The Heavy Flavor Averaging Group (HFAG)~\cite{HFAG} has recently
performed a first comprehensive study of $|V_{\rm ub}|$ from inclusive decay
measurements using DGE. This analysis, which is summarized in
Fig.~\ref{HFAG_Vub}, provides
the most precise determination of $|V_{\rm ub}|$ so far.
Since different measurements in Fig.~\ref{HFAG_Vub}
correspond to different kinematic cuts with
extrapolation factors ranging between $\sim 2$ and~$5$,
the consistency of the resulting values for
$|V_{\rm ub}|$ provides an additional evidence that the underlying description
of the spectrum is good. Direct comparison of the DGE calculation\footnote{The DGE calculation has
been implemented numerically in \texttt{C++} facilitating
phase--space integration with a variety of cuts. The program is
available at {\tt
http://www.hep.phy.cam.ac.uk/$\sim$andersen/BDK/B2U/}. } of the
triple differential
$\bar{B}\longrightarrow X_u l \bar{\nu}$ spectrum~\cite{Andersen:2005mj}
(or its moments with varying cuts) with
experimental data would be useful to constrain power corrections. Such analysis is
complementary to the $\bar{B}\longrightarrow X_s \gamma$ one in several ways
owing to the potential Weak Annihilation contributions and
the different (non-universal) subleading power
corrections in the two
processes~\cite{Bauer:2002yu}--\cite{Lee:2004ja},
effects that are hard to quantify theoretically and that
have so far been neglected.

Having seen the advancement achieved by DGE in the calculation of inclusive
decay spectra, let us now return to
the theoretical foundation of the approach. Infrared and collinear
safe observables, such as inclusive decay spectra or event--shape distributions,
typically involve two distinct sources of large corrections, the
jet function and
the soft function in \eq{factorization}. We begin in Sec.~\ref{DIS}
by considering the example of the
jet function $J_N(Q,\mu_F)$,
which is common to a large class of inclusive cross sections and decay
spectra~\cite{Gardi:2001di,CG,BDK}. In Sec.~\ref{SOFT} we turn to
the soft function $S_N(Q,\mu_F)$.

\section{An unresolved jet and deep inelastic structure functions near the
elastic limit\label{DIS}}

Sudakov resummation is a manifestation of the infrared safety of
inclusive distributions at the logarithmic level. However, the
Sudakov evolution kernel conceals infrared sensitivity at the power
level, which only becomes explicit once running--coupling effects
are resummed to all orders. Consequently, \emph{Sudakov resummation
involves power corrections that exponentiate along with the
logarithms.}

To demonstrate this, consider first the jet function
\cite{Sterman:1986aj,CT} $J_N(Q;\mu_F)$ that can be defined in a
process independent way as the quark propagator in axial gauge (Eq.
(2) in~\cite{GR}). It obeys the following evolution equation:
\begin{eqnarray}
\label{J_ev}
\frac{d\ln J_N(Q;\mu_F)}{d\ln Q^2}= \int_0^1 dx\frac{ x^{N-1}-1}{1-x} {\cal
J}\left(\alpha_s((1-x)Q^2)\right).
\end{eqnarray}
${\cal J}\left(\alpha_s\right)$ is the corresponding {\em
scheme--invariant} anomalous dimension. It is usually decomposed
in the ${\overline{\rm MS}}$ scheme as~\cite{GR}
\begin{eqnarray}
\label{J} {\cal J}\left(\alpha_s(\mu^2)\right)={\cal
A}\left(\alpha_s(\mu^2)\right)+ \frac{d\alpha_s(\mu^2)}{\ln\mu^2}
\frac{d{\cal B}
\left(\alpha_s(\mu^2)\right)}{d\alpha_s}=C_F\frac{\alpha_s(\mu^2)}{\pi}+
\ldots,
\end{eqnarray}
where the conformal part, ${\cal A}\left(\alpha_s\right)$, is the
universal cusp anomalous dimension, which is the coefficient of the
singular part, $1/(1-x)_+$, in the splitting
function~\cite{Korchemsky:1988si}. ${\cal A}$ alone controls
the factorization scale dependence of the jet function:
\begin{eqnarray}
\label{DIS_muF}
 \frac{d\ln
J_N(Q;\mu_F)}{d\ln \mu_F^2}=
{\cal A}\left(\alpha_s(\mu_F^2)\right)\ln N.
\end{eqnarray}
The anomalous dimension ${\cal B}(\alpha_s)=-\frac34\frac{\alpha_s}{\pi}+\ldots$
is associated with collinear singularities~\cite{Sterman:1986aj,CT}
in the axial--gauge light--quark propagator, and its perturbative
expansion is currently known to NNLO~\cite{MVV,Moch:2005ba}.

The jet function $J_N(Q;\mu_F)$ is also the Sudakov factor controlling the
$x\longrightarrow 1$ limit of deep inelastic
structure functions~\cite{GR},
\begin{equation}
\label{F2}
F_2^N(Q^2)\equiv \int_0^1 dx x^{N-2} F_2(x,Q^2)
\simeq \underbrace{\Big[H(Q;\mu_F) \, J_N(Q;\mu_F) \,+\,{\cal O}(1/N)\Big]}_{\rm
 twist-two\,\, coefficient\,\, function}
\,\times\, q_N(\mu_F)  \,+\,
{\cal O}(N\Lambda^2/Q^2),
\end{equation}
where $q_N(\mu_F)$ is the quark distribution function,  and
$H(Q;\mu_F)$ is a hard function that does not depend on $N$.
The evolution of the quark distribution function
$q_N(\mu_F)$ is controlled at large $N$
by the cusp anomalous dimension ${\cal A}\left(\alpha_s\right)$:
\begin{eqnarray}
\label{q_muF}
 \frac{d\ln q_N(\mu_F)}{d\ln\mu_F^2}=
-{\cal A}\left(\alpha_s(\mu_F^2)\right)\ln N\,+\,{\cal O}(1)
\end{eqnarray}
such that the product in \eq{F2} is invariant.

The evolution equation of the structure function itself (at leading twist) is
\begin{eqnarray}
\label{DIS_physical}
 \frac{d\ln F_2^N(Q^2)}{d\ln Q^2}&=&\frac{d\ln
J_N(Q;\mu_F)}{d\ln Q^2}+\frac{d\ln H(Q;\mu_F)}{d\ln Q^2}+{\cal O}(1/N)\\
&=& \int_0^1 dx\frac{ x^{N-1}-1}{1-x} {\cal\nonumber
J}\left(\alpha_s((1-x)Q^2)\right)+\frac{d\ln H(Q;\mu_F)}{d\ln Q^2}+{\cal O}(1/N),
\end{eqnarray}
implying that log--enhanced terms in the leading--twist
coefficient function of $F_2$
exponentiate in moment space, to any order in perturbation theory.

The cancellation between real $(x^{N-1})$
and virtual $(-1)$ contributions guaranties that
the evolution kernel on
 the r.h.s of \eq{DIS_physical} or \eq{J_ev}
is finite at any order in perturbation theory, despite the $1/(1-x)$
singularity.
Given that the anomalous dimension ${\cal
J}\left(\alpha_s(\mu^2)\right)$ is known to a certain order
(currently NNLO~\cite{MVV}, with a good estimate of the
N$^3$LO~\cite{Moch:2005ba}) one can use the $\beta$ function and integrate
the r.h.s of Eq.~(\ref{J_ev}) order by order, obtaining
the following expansion:
\begin{equation}
{\rm r.h.s\, \Big\{\eq{J_ev}
\Big\}
}\,=\,C_F\,
\sum_{j=1}^{\infty}\sum_{l=1}^{j}
c_{j,l}\,\left(\frac{\alpha_s(Q^2)}{\pi}\right)^j (\ln N)^l,
\label{kernel_sum_}
\end{equation}
where the coefficients $c_{j,l}$ with $l=j$ at any order $j$
(leading logarithms, LL) are determined by the LO term in ${\cal
J}\left(\alpha_s\right)$, those with $l=j-1$ (next--to--leading
logarithms, NLL) require also the NLO term in ${\cal
J}\left(\alpha_s\right)$, etc.
The standard approach to Sudakov resummation is based on expressing the sum
in \eq{kernel_sum_} as an expansion with increasing {\em logarithmic accuracy}, namely
\begin{equation}
{\rm r.h.s\, \Big\{\eq{J_ev}
\Big\}
}\,=\,\frac{C_F}{\beta_0}\,
\sum_{n=0}^{\infty} g_n(\lambda)
\left(\frac{\alpha_s(Q^2)}{\pi}\right)^n\,;\qquad\quad
\lambda \equiv \frac{\beta_0\alpha_s(Q^2)}{\pi}\,\ln N,
\label{kernel_sum_1}
\end{equation}
where $\beta_0=\frac{11}{12}C_A-\frac{1}{6}N_f$.
Here\footnote{A well--known fact is that $g_n(\lambda)$ have Landau singularities
at $\lambda=1$, corresponding to $Q^2/N=\Lambda^2$. Obviously, the perturbative
analysis is valid only when the jet--mass scale $Q^2/N$ is sufficiently large
compared to $\Lambda^2$. However, when inverting the Mellin transform
according to \eq{inv_Mellin} one needs to deal with this spurious singularity
for any value of $x$~\cite{Catani:1996dj}. This problem is completely
avoided in the DGE approach.} $g_0(\lambda)=\ln(1-\lambda)$, and higher--order
terms $g_n(\lambda)$ are determined based on the expansion of
${\cal J}\left(\alpha_s\right)$ to N$^n$LO.
This sum is therefore truncated at $n=n_{\max}$, the order at which ${\cal
J}\left(\alpha_s\right)$ has been computed.

The next, crucial observation~\cite{DGE_thrust,Gardi:2001di,GR} is  that
the sum in \eq{kernel_sum_1}
does not converge. The same is true for any reorganization of the terms in
\eq{kernel_sum_} because the coefficients $c_{j,l}$ increase as $j!$ at
high orders. This is the effect of an infrared renormalon that dominates
the evolution kernel in \eq{J_ev}
at high orders. Importantly, ${\cal J}$,  being an anomalous dimension,
 is expected to be \emph{free of any renormalon
singularities}\footnote{This
is a conjecture. Explicit calculation
in the large--$\beta_0$ limit (\eq{large_b0_J} below) supports it.}.
Thus, the one and only source
of factorial divergence in \eq{J_ev} is the integration  over the momentum fraction $x$ near
$x\longrightarrow 1$. In this limit the coupling in ${\cal
J}\left(\alpha_s(\mu^2)\right)$  is probed at extremely soft momentum scales
$\mu^2\longrightarrow 0$. The non-existence of the sum in Eqs.
(\ref{kernel_sum_1}) or (\ref{kernel_sum_}) reflects
sensitivity to the far infrared at the power level.

A systematic way to quantify this infrared sensitivity is to
regularize the divergence as a Borel sum. To this end, imagine that
the anomalous dimension ${\cal J}\left(\alpha_s\right)$ on the r.h.s. of
\eq{J_ev} were known to all orders, so one could write
its scheme--invariant Borel representation:
\begin{equation}
\label{Borel_rep_J} {\cal
J}\left(\alpha_s(\mu^2)\right)=\frac{C_F}{\beta_0} \int_0^{\infty}du
\left(\frac{\Lambda^2}{\mu^2}\right)^u T(u)\, B_{\cal J}(u),
\end{equation}
where $T(u)$ is the Borel transform of the coupling\footnote{In general we
use the scheme--invariant Borel transform~\cite{Grunberg:1992hf} where
$T(u)$ is the Borel transform of the
two-loop coupling, see Eq. (2.18) in Ref.~\cite{Andersen:2005bj}.
In the large--$\beta_0$ limit $T(u)=1$.}.
Using~Eq.~(\ref{Borel_rep_J}) in \eq{J_ev} (or in \eq{DIS_physical}) one can
perform the $x$ integration arriving at:
\begin{eqnarray}
\label{J_ev_Borel} &&\hspace*{-20pt}
\frac{d\ln
J_N(Q;\mu_F)}{d\ln Q^2}= \frac{C_F}{\beta_0}\int_0^{\infty} du
\left(\frac{\Lambda^2}{Q^2}\right)^u T(u)  \, B_{\cal
J}(u)\,\left[\frac{\Gamma(N)\Gamma(-u)}{\Gamma(N-u)}+\frac{1}{u}
\right].
\end{eqnarray}
Indeed, one finds potential renormalon ambiguities at positive
integer values of $u$ arising from the $x\longrightarrow 1$ limit.
As mentioned above, $B_{\cal J}(u)$ is not
expected to have any renormalon singularities, however, unless it
\emph{vanishes} at $u=k$ where $k=1,2,\ldots$, the evolution kernel
(the r.h.s. of \eq{J_ev_Borel}) will have
\emph{power-like ambiguities} that scale at large $N$ as
$(N\Lambda^2/Q^2)^k$. Upon solving the evolution equation for $J_N(Q;\mu_F)$,
any such power terms would obviously exponentiate
along with the logarithms.

The discussion of renormalons can be made concrete by focusing on
the gauge--invariant
 set of radiative corrections corresponding to the
 large--$\beta_0$ limit\footnote{Results in the large--$\beta_0$ limit are
obtained~\cite{Beneke:1998ui} by first considering the large--$N_f$
limit, in which a gluon is dressed by any number of fermion--loop
insertions, and then making the formal substitution
$N_f\longrightarrow -6 \beta_0$. }. In this limit one can get
analytic results for the Borel transform. The result for the
anomalous dimension reads~\cite{GR}:
\begin{equation}
\label{large_b0_J} B_{\cal J}(u)\bigg\vert_{{\rm
large}\,\beta_0} =\,\frac{{\rm e}^{\frac53 u}}{2}\,
\frac{\sin\pi u}{\pi u} \,\left(\frac{1}{1-u}+\frac{1}{1-u/2}\right),
\end{equation}
implying that renormalon singularities in \eq{J_ev_Borel}
at $u=1$ and $u=2$ are indeed
present\footnote{Eq.~(\ref{large_b0_J}) is consistent with
renormalon calculations of deep inelastic coefficient functions, done
independently of the $x\to 1$ limit~\cite{Dasgupta:1996hh,Stein:1996wk}.}.
Consequently, Sudakov
resummation cannot be considered a purely perturbative issue: the
evolution kernel itself is only defined in perturbation theory up to
powers. Since the l.h.s of Eq.~(\ref{DIS_physical}) is an
observable, these ambiguities must cancel once non-perturbative
effects are systematically included. Indeed, within the OPE renormalon
ambiguities reflect the mixing under
renormalization between operators of different
twist~\cite{Beneke:1998ui,Gardi:2002bk,Braun:2004bu}; in the case of $F_2$
this issue is understood in full detail at twist four, see
Ref.~\cite{Gardi:2002bk}. Of course, the OPE cannot be directly applied to the
evolution equation (\ref{J_ev_Borel}), which was only derived at the
leading--twist level. The \emph{exponentiation} of these power
corrections~\cite{Gardi:2002bk,GR}
goes beyond the OPE analysis: it amounts to resumming the OPE.

Importantly, $B_{\cal J}(u)$ in \eq{large_b0_J} vanishes at certain
integer values $u=k$, for $k\geq 3$, so the corresponding renormalon
ambiguities in the evolution kernel (\ref{J_ev_Borel}) are absent.
Since a single dressed gluon is a natural approximation to the
evolution kernel (e.g. iteration of a single chain is generated by
solving the equation, through
exponentiation) we expect that
the pattern of renormalon singularities exposed in the
large--$\beta_0$ limit would not be modified. We will therefore
assume that the zeros of $B_{\cal J}(u)$ in \eq{large_b0_J}
\emph{are} the zeros of this function in the full theory. As we
shall see in the next section, other Sudakov anomalous dimensions
are characterized by their own pattern of zeros.

In general, Borel singularities in QCD differ from their
large--$\beta_0$ limit in both the value
of the residue and
the nature of the singularity~\cite{Mueller:1984vh,Beneke:1998ui}.
However, Borel singularities in the Sudakov evolution kernel
remain\footnote{Note that these are simple poles only in the
scheme--invariant formulation of the Borel transform, where the Borel
variable is conjugate to $\ln(Q^2/\Lambda^2)$. In the standard formulation,
where the Borel variable is conjugate to $\pi/(\beta_0\alpha_s)$ the
singularities turn into cuts, controlled by $\beta_1/\beta_0^2$. See
Eq. (B.13) in Ref.~\cite{Andersen:2005bj}.} \emph{simple poles}\footnote{It implies~\cite{Gardi:2002bk},
in particular, that the mixing of higher--twist operators into the leading
twist involves at large $N$ pure powers, with no additional logarithms.
In other words, one expects that the effective anomalous dimensions
of higher--twist operators would have the \emph{same}
large--$N$ asymptotic behavior as the leading--twist operator,
\eq{q_muF}.\\
Anomalous dimensions of higher--twist operators can of course be computed
directly, independently of their power--like mixing with the leading twist.
A significant progress in addressing this problem
was made in recent years using
the integrability property of lightcone
operators~\cite{Belitsky:2004cz}.
A general result is that the spectrum of higher--twist
anomalous dimensions, computed as a function of $N$,
is bounded from below by the leading--twist anomalous dimension --- see e.g.
Eqs. (4.4) and (5.3) in Ref.~\cite{Braun:2001qx}.
This implies that the above expectation is indeed realized
asymptotically.}~\cite{GR}. This follows directly
from \eq{J_ev_Borel} and the above conjecture that $B_{\cal J}(u)$
itself does not have Borel singularities.

To make practical use of Eq.~(\ref{J_ev_Borel}) one obviously needs
sufficient knowledge of the Borel transform of the
Sudakov anomalous dimension, $B_{\cal J}(u)$. Perturbation theory directly
 gives the Taylor expansion of $B_{\cal J}(u)$ near the origin.
Examining the integral in \eq{J_ev_Borel} one finds that
at small $N$ only the immediate vicinity of the origin
is relevant, while for $N \sim {\cal O}(Q^2/\Lambda^2)$ contributions from
$u\sim 1$ and $u\sim 2$ become important.
Owing to the strong suppression of contributions from large $u$
through $\Gamma(-u)$, the behavior of $B_{\cal J}(u)$ at
large $u$ is irrelevant, even for high moments.
Aiming at power accuracy, one would need to have
$B_{\cal J}(u)$ under control at least for $0\leq u\lsim  1$, and specifically,
constrain the value of $B_{\cal J}(u=1)$ that controls the magnitude of the
leading renormalon residue in the evolution kernel.

It is natural to express $B_{\cal J}(u)$ in terms of its
known large--$\beta_0$ limit
as follows\footnote{Starting with the NNLO result for $B_{\cal J}(u)$,
Ref.~\cite{GR} considered several different functional
forms that are consistent with \eq{large_b0_J},
concluding that a multiplicative form as in \eq{BJ_full},
where $V_{\cal J}(u)$ does not introduce any Borel singularities,
results in good apparent convergence of the
logarithmic accuracy expansion, whereas \eq{large_b0_J} with
additive ${\cal O}(1/\beta_0)$
corrections would lead to extremely large subleading logarithms.
Ref.~\cite{Moch:2005ba} has shown that the logarithmic accuracy expansion
converges well. This supports the multiplicative form of \eq{BJ_full}.
The new N$^3$LO results~\cite{Moch:2005ba} constrain $B_{\cal J}(u)$
further, and they may well facilitate an approximate
determination of $B_{\cal J}(u=1)$ corresponding to the leading renormalon
residue in~\eq{J_ev_Borel}.}:
\begin{equation}
\label{BJ_full}
B_{\cal J}(u)=\left. B_{\cal
J}(u)\right\vert_{{\rm large }\,\beta_0}\times V_{\cal J}(u)
\,;\qquad\qquad V_{\cal J}(u)=1+{\cal
O}(u/\beta_0),
\end{equation}
where $V_{\cal J}(u)$ accounts for contributions that are subleading for
$\beta_0\longrightarrow \infty$; its Taylor expansion
around $u=0$ can be determined order by order in perturbation theory.
\eq{BJ_full} is consistent with $B_{\cal J}(u)$ having the same
pattern of zeros as exposed by the large--$\beta_0$ limit
(\ref{large_b0_J}) provided that $V_{\cal J}(u)$ itself has no
singularities nor zeros, at least for positive integer values of $u$.
A concrete ansatz\footnote{When using \eq{BJ_full} for phenomenology,
the sensitivity of the result to the particular ansatz for $V_{\cal J}(u)$
represents the theoretical uncertainty due to unknown higher--order corrections.}
that satisfies these
requirements is $V_{\cal J}(u)=\exp (-w_1 u-w_2 u^2+\cdots)$.
Here the coefficients $w_i$ are determined order by order,
up to the order at which ${\cal J}(\alpha_s)$ has been
computed~\cite{GR}:
based on the NLO result for the cusp anomalous dimension and \eq{J}
one finds
$w_1=\frac{C_A}{\beta_0}\left(\frac{\pi^2}{12}-\frac13 \right)$,
from the NNLO result for ${\cal J}(\alpha_s)$ one can determine $w_2$, see
Eq. (2.33) in \cite{Andersen:2005bj}, and finally from the
approximate~\cite{Moch:2005ba} N$^3$LO coefficient for ${\cal J}(\alpha_s)$
one can determine $w_3$ to good accuracy.

The characteristic feature of the power ambiguities of
Eq.~(\ref{J_ev_Borel}) is their enhancement by the moment
index $N$, reflecting the fact that the physical scale involved is
$Q^2/N$, the invariant mass of the jet. In the
language of non-local lightcone operators~\cite{BB}, the large--$N$
limit corresponds to the limit of large lightcone separation between the quark
fields~\cite{Gardi:2002bk}. Alternatively, within the local OPE this
parametric enhancement is understood as power increase in the
\emph{number} of relevant local operators. This leads to an explosion
in the number of non-perturbative parameters, making the description
of structure functions in the large--$x$ limit within the
framework of the OPE completely impractical.

Physically, the breakdown of the OPE and the appearance of
non-perturbative corrections on the jet mass scale are related to
the transition into the resonance
region~\cite{Liuti:2001qk,Melnitchouk:2005zr}, where eventually the
jet becomes a few exclusive states. Obviously, the wealth of
non-perturbative phenomenon in this region cannot be described in
detail using the OPE.
However, parton--hadron duality may well work in moment space
up to very large $N$.

The discussion above suggests a clear track
along which the evolution of structure functions at large--$x$
should be computed~\cite{Gardi:2002bk,GR}:
\begin{itemize}
\item{} First, the evolution kernel on the r.h.s of
\eq{J_ev_Borel} needs to be regularized, e.g. by taking the
Principal Value of the Borel sum.
Upon solving the evolution equation with the regularized kernel (and matching
with the fixed--order result)
one obtains an improved approximation to the evolution of the
structure function at large~$N$, which is valid up to ${\cal O}(N\Lambda^2/Q^2)$
corrections.
\item{} For sufficiently large~$N$, the
${\cal O}(N\Lambda^2/Q^2)$ power ambiguities become relevant, and
 the perturbative evolution
should be supplemented by non-perturbative power corrections.
Since power ambiguities appear in the evolution kernel,
power corrections appear in the Sudakov {\em exponent}. The
perturbative jet function is then modified by a \emph{multiplicative}
non-perturbative factor of \emph{a particular
 form} that is dictated by the
 ambiguities\footnote{Note that although the analysis is formally
valid up to $1/N$ terms, we keep the full $N$ dependence of the ambiguities. This is
a convenient choice; for example it implies that the normalization
convention of $J_N^{\rm PV}(Q,\mu_F)$ (vanishing at $N=1$) are
then satisfied by $J_N^{\rm NP(PV)}(Q)$.},
\begin{eqnarray}
\label{J_NP}
&&J_N(Q,\mu_F)\,\longrightarrow\, J_N^{\rm PV}(Q,\mu_F)\,\times\,J_N^{\rm NP(PV)}
\left(Q\right);\nonumber \\\nonumber \\
&&{\rm where} \qquad J_N^{\rm NP(PV)}\left(Q\right)=\exp\Big({
-\omega_1^{\rm PV}(N-1) \left({\Lambda^2}/{Q^2}\right)
-\omega_2^{\rm PV}(N-1)(N-2)\left({\Lambda^2}/{Q^2}\right)^2}\Big).
\end{eqnarray}
The non-perturbative
parameters $\omega_{1,2}^{\rm PV}$ correspond to the regularization prescription
(Principal Value) applied to the r.h.s of \eq{J_ev_Borel}, such that
the product $J_N^{\rm PV}(Q,\mu_F)\,\times\,J_N^{\rm NP(PV)}(Q)$
is prescription--independent.
It should be emphasized that by introducing power terms
according to the ambiguity structure we
explicitly assume a {\em minimal model} for the non-perturbative
contribution: the first two power terms must definitely be there, but
it is impossible to \emph{exclude} by present theoretical tools
higher power terms.
To the extent that this assumption holds, \eq{J_NP}
sums up the dominant corrections in the OPE,
scaling as powers of $N\Lambda^2/Q^2$, to all orders
({\it cf.} \eq{F2})~\cite{Gardi:2002bk}:
\begin{equation}
\label{F2_NP_factorization}
F_2^N(Q^2)
\,\simeq\,\Big[ H(Q;\mu_F) \, J_N^{\rm PV}(Q;\mu_F)\,\times
\,J_N^{\rm NP(PV)}\left(Q\right)\,+\,{\cal O}(1/N)\Big]
\,\times\, q_N(\mu_F) \,+\,
{\cal O}(\Lambda^2/Q^2).
\end{equation}
\end{itemize}

Ultimately, large--$x$ structure function data can
be analyzed along these lines. Conventional structure function
phenomenology~\cite{Martin:2003sk} (see,
however, Refs.~\cite{CM,GR,Yang:1999xg,Alekhin:1998df}) does
not focus on constraining
high moments, and parton--distribution--function fits are typically
done with a cut on the hadronic mass,
$Q^2 (1-x)/x > 12.5\, {\rm GeV}^2$, restricting the
data considered. In this way one avoids the need for
Sudakov resummation or higher--twist corrections.
Obviously, such restrictions limit the possibility of constraining
the parton distribution functions.
With the theoretical tools described above,
there are high prospects for improving these constraints.

\section{Soft gluons, Sudakov anomalous dimensions and parametrically--enhanced
power corrections\label{SOFT}}

In the previous section we discussed the DGE approach in the context of the
exclusive limit of deep inelastic structure functions.
This problem is special since it is characterized by a single hierarchy: the
invariant mass of the jet~$Q^2(1-x)/x$ gets significantly smaller than the hard
momentum transfer $Q^2$;
when the jet--mass scale approaches the hadronic scale, the
structure function directly probes the exclusive limit,
where the jet is composed of a few hadrons.
As explained in the introduction, other inclusive
distributions
have a more complex structure, where the
threshold region is characterized by a double
hierarchy. Specifically,
infrared and collinear safe observables such as
event--shape distributions and inclusive decay spectra involve the following
three scales:
hard ${\cal O}(Q)$, ``jet''  ${\cal O}(Q/\sqrt{N})$
and ``soft''  ${\cal O}(Q/N)$. In this situation the threshold region,
where Sudakov resummation is important, is parametrically wider than the
exclusive region discussed above. Moreover,
the most important power corrections are then the ones
associated with large--angle soft emission, powers of $N\Lambda/Q$, whereas
the non-perturbative structure of the jet, i.e.
powers of $N\Lambda^2/Q^2$, can usually be neglected.

\begin{figure}[t]
\includegraphics[width=125mm]{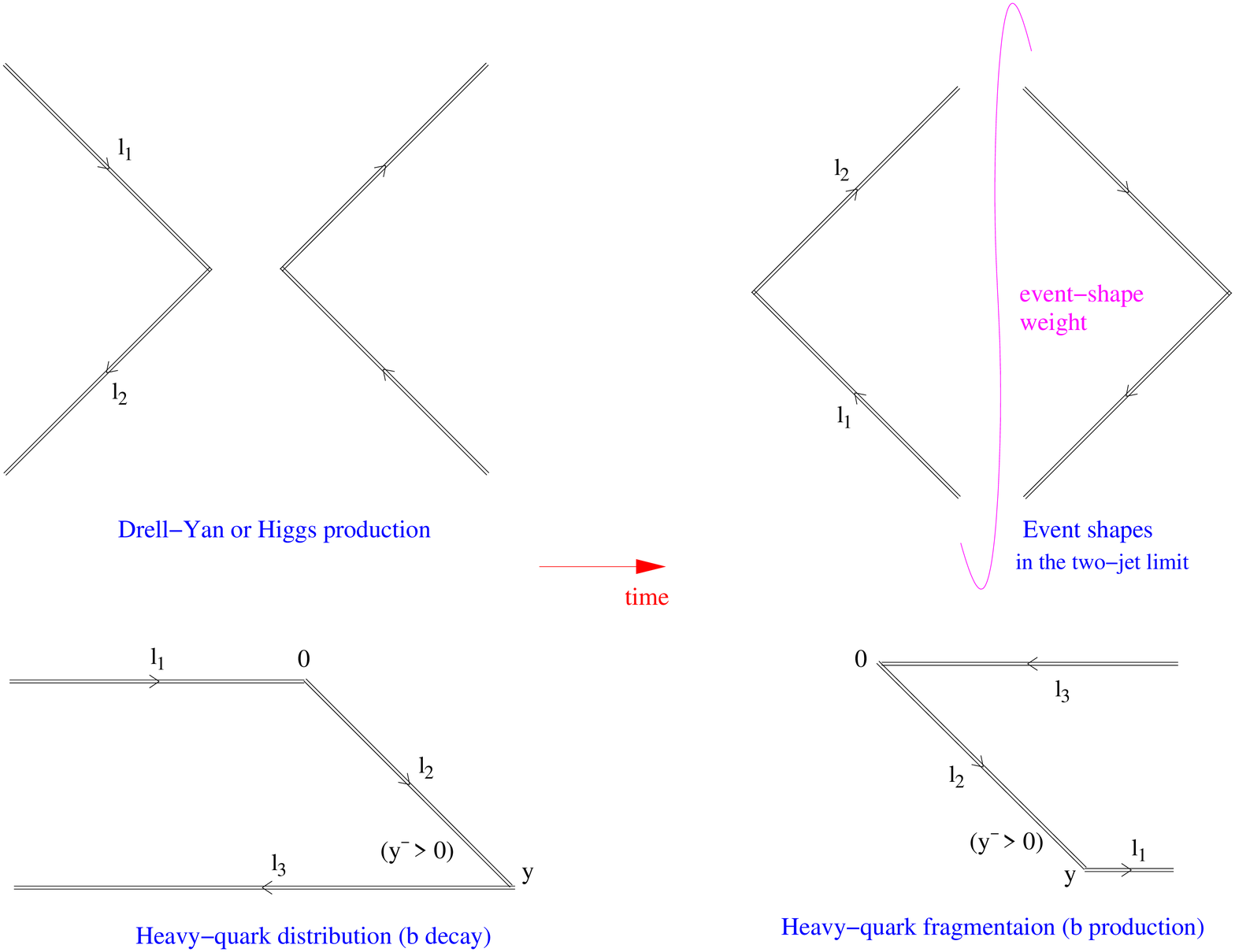}
\caption{Representation of the soft Sudakov factor $S_N(Q;\mu_F)$ in
terms of Wilson--line operators in Minkowski space--time, where the
time is the horizontal axis. The four diagrams correspond to
Drell--Yan or Higgs production, event--shape distributions in the
two--jet limit, heavy--quark distribution (inclusive decays), and
heavy--quark fragmentation (heavy--meson production). In each case
the lines close at infinity to form a loop, which guarantees gauge
invariance.
 \label{Wilson}}
\end{figure}

In this section we consider Sudakov factors associated with
soft--gluon emission, which we denote in general by $S_N(Q;\mu_F)$.
The  Sudakov factor $S_N(Q;\mu_F)$ is defined at the level of the
cross section so it involves the amplitude times the complex
conjugate amplitude.
Soft gluons do not resolve the hard interaction nor the structure of the jet;
they couple to the color field along the
classical trajectories of hard
partons~\cite{Gatheral:1983cz,Frenkel:1984pz,Sterman:1986aj}.
The relevant
log--enhanced terms can therefore be computed in the Eikonal approximation
or, equivalently, by considering Wilson--line
operators~\cite{KR87,Korchemsky:1991zp,Korchemsky:1988si,Korchemsky:1992xv,Korchemskaya:1992je} ---
see Fig.~\ref{Wilson}.
Such field--theoretic definitions are useful because they readily generalize
the perturbative Sudakov factor to the non-pertubative level~\cite{Korchemsky:1992xv,QD,Korchemsky:1999kt}.

``Soft'' Sudakov factors corresponding to different observables are
the same \emph{only} in the conformal limit (or
to leading logarithmic accuracy) where just the cusps along the Wilson
loop contribute; subleaing logarithms and
power corrections depend on the \emph{geometry} of the process
(Fig.~\ref{Wilson}) and
on the way particle momenta are \emph{weighted}.
Indeed, already at the perturbative level
$S_N(Q;\mu_F)$ have different physical content
in different processes, for example:
\begin{itemize}
\item{} In Drell--Yan or Higgs production near partonic\footnote{If the initial--state
parton distribution functions are defined in
dimentional--regularization based factorization schemes (such as ${\overline{\rm MS}}$),
these cross sections involves \emph{only} a ``soft''
Sudakov factor~\cite{Sterman:1986aj,CT,CMW}; a jet--mass scale never appears.} threshold
$S_N(Q;\mu_F)$ describes the \emph{energy distribution}
of soft gluons ($1-x$ is the small energy fraction carried by gluons)
emitted by the incoming partons,
which can be represented in terms of lightlike Wilson
 lines~\cite{CMW,Contopanagos:1996nh,Sterman:1986aj,Collins:1984kg,Korchemsky:1994is,Beneke:1995pq,Gardi:2001di,Laenen:2000ij}.
\item{} In a large class\footnote{We consider
observables where contributions from individual soft gluon
emissions are additive, leading to exponentiation in moment space.}
of event--shape distributions in $e^+e^-$
annihilation~\cite{DGE_thrust,Gardi:2003iv,Gardi:2002bg}, $S_N(Q;\mu_F)$
 stands for the
contribution to the observable by
(large--angle) soft gluons that are emitted  off the recoiling
quark--antiquark in the two--jet limit. It
can formally be described in terms of a weighted integral over a matrix element
involving lightlike Wilson lines~\cite{Korchemsky:1999kt,Belitsky:2001ij}.
\item{} In heavy--quark fragmentation $S_N(m;\mu_F)$ describes soft gluon
radiation off an off--shell heavy quark that was initially
produced with high energy of ${\cal O}(Q)$, where $Q\gg m$,
and then gradually looses its energy approaching its mass shell.
This radiation is dominated by
transverse momenta (or virtualities) of
order \hbox{$|k_{\perp}|\sim m(1-x)$},
--- the ``soft'' scale ---
where $x$ is the longitudinal momentum fraction carried by
the final--state on-shell heavy quark
(which represents the detected heavy hadron
on the perturbative level)~\cite{DKT,DKT95,CC,CG,Gardi:2003ar,Melnikov:2004bm}. The description of $S_N(m;\mu_F)$
in terms of Wilson lines involves timelike lines connected by a
finite lightlike segment
of length $y^-$ that is proportional to $N$~\cite{Korchemsky:1992xv,QD}.
\item{} In heavy--meson decay spectra $S_N(m;\mu_F)$ describes the soft gluon
interaction with the heavy quark prior to its decay. Formally,
$S_N(m;\mu_F)$
is the Sudakov factor associated with the
longitudinal momentum distribution of an off-shell heavy--quark field
in an on-shell heavy--quark initial state. Remarkably, it is \emph{identical} to the
Sudakov factor in heavy--quark fragmentation~\cite{QD}, while
the non-perturbative dynamics in the two cases is different.
\end{itemize}

Despite the different physics described by the ``soft'' Sudakov factor
in different processes, $S_N(Q;\mu_F)$ has a universal structure
that is constrained by factorization, infrared safety and
renormalization--group invariance. In particular, in all the examples
mentioned above, upon neglecting ${\cal O}(1/N)$ corrections
one can establish an evolution equation that
is analogous to the jet--function
case~(\ref{J_ev}):
\begin{eqnarray}
\label{Soft} \frac{d\ln S_N(Q;\mu_F)}{d\ln Q^2}\,=\,-\int_0^1 dx\, \frac{
x^{N-1}-1}{1-x} \,\,{\cal S}\left(\alpha_s((1-x)^2Q^2)\right).
\end{eqnarray}
This equation implies exponentiation of all the logarithms to any
order in perturbation theory. The \emph{observable--dependent}, but
\emph{scheme--invariant} anomalous dimension ${\cal
S}\left(\alpha_s\right)$ is usually decomposed in ${\overline {\rm
MS}}$ as:
\begin{eqnarray}
\label{S} {\cal S}\left(\alpha_s\right)={\cal
A}\left(\alpha_s\right)\!+\! \frac{d\alpha_s}{\ln\mu^2} \frac{d{\cal
D}\left(\alpha_s\right)}{d\alpha_s}=C_F\frac{\alpha_s}{\pi}+\ldots,
\end{eqnarray}
where ${\cal A}(\alpha_s)$ is the universal cusp anomalous dimension appearing
in \eq{J} and ${\cal D}(\alpha_s)$ is observable--dependent. The
dependence of $S_N(Q;\mu_F)$ on $\mu_F$ is the same as that of the
$\overline{\rm MS}$ quark distribution, i.e. it has the opposite
sign to Eq.~(\ref{DIS_muF}) above, making the product in
\eq{factorization} factorization--scale invariant.

Similarly to the jet--function case, infrared sensitivity at power level is
concealed in the r.h.s to the evolution equation~(\ref{Soft}): in
the $x\longrightarrow 1$ limit the coupling is probed at extremely
soft momentum
scales, $\mu^2=Q^2 (1-x)^2 \longrightarrow 0$. In conventional Sudakov resummation this
sensitivity is not regularized and the resulting divergent series is
truncated at some order (some logarithmic accuracy).
In order to regularize the divergence and quantify the
infrared sensitivity as power--suppressed ambiguities we introduce a Borel
representation\footnote{As before we extract the color factor out of the
Borel transform. $C_F$ in \eq{Borel_rep_S} refers to radiation off hard
\emph{quarks}; For Higgs production though gluon--gluon fusion it
should be replaced by $C_A$. This issue aside, $S_N(Q;\mu_F)$
is identical to the Drell-Yan case.} for the anomalous dimension
${\cal S}\left(\alpha_s\right)$ in analogy
with \eq{Borel_rep_J}:
\begin{equation}
\label{Borel_rep_S} {\cal
S}\left(\alpha_s(\mu^2)\right)=\frac{C_F}{\beta_0} \int_0^{\infty}du
\left(\frac{\Lambda^2}{\mu^2}\right)^u T(u) B_{\cal S}(u).
\end{equation}
Using (\ref{Borel_rep_S}) in (\ref{Soft}), and performing the $x$
integration, one obtains the DGE form of the evolution
equation:
\begin{eqnarray}
\label{Soft_Borel}  \frac{d\ln S_N(Q;\mu_F)}{d\ln Q^2}=-
\frac{C_F}{\beta_0}\int_0^{\infty} du \left(\frac{\Lambda^2}{Q^2}
\right)^u  T(u)  \times B_{\cal S}(u)
\times\left[\frac{\Gamma(N)\Gamma(-2u)}{\Gamma(N-2u)}+\frac{1}{2u}
\right].
\end{eqnarray}
Note that while Eq.~(\ref{J_ev_Borel}) indicates potential
renormalon singularities at all {\em integer} values of $u$, in
Eq.~(\ref{Soft_Borel}) renormalons may occur also at {\em half
integer} values. The corresponding power ambiguities at $u=\frac{k}{2}$
($k$ being a positive integer)
would scale at large $N$ as $(N\Lambda/Q)^k$.
Specifically, the leading renormalon ambiguity can
be at $u=\frac12$, which would result in a substantial ${\cal
O}(N\Lambda/Q)$ effect
--- a global shift of the distribution. Consequently, these power
corrections received much attention in various applications
including Drell--Yan production
\cite{Korchemsky:1994is,Beneke:1995pq,Gardi:2001di} (where the
$u=\frac12$ ambiguity was found \emph{not} to occur, see below),
heavy--quark fragmentation~\cite{Nason:1996pk,CG,Gardi:2003ar} and
especially event--shape
variables~\cite{Korchemsky:1995zm}--\cite{BM}.
Notably, the relevance of renormalons in computing inclusive B decay
spectra has not received proper attention\footnote{See, however,
earlier work in Ref.~\cite{Grozin:1994ni}.} until
recently~\cite{BDK,QD,Andersen:2005bj,Andersen:2005mj,Gardi:2004gj,Gardi:2005mf}.

Since ${\cal S}\left(\alpha_s\right)$ is an anomalous dimension,
 $B_{\cal S}(u)$ is not expected to have any renormalon singularities
 of its own.
It follows that the sole origin of renormalons in the evolution kernel is the
integration over $x$ for $x\longrightarrow 1$, leading to
 {\em simple poles} in Eq.~(\ref{Soft_Borel})\footnote{Going beyond
 the large--$\beta_0$ limit
the singularities remain simple poles only
 upon using the scheme--invariant formulation of the
 Borel transform (\ref{Soft_Borel}),
 where the effect of the two-loop running coupling is
 factored out into $T(u)$.
In a standard Borel representation with respect to the coupling, the
simple poles will be replaced by cuts whose strength is determined
by the first two coefficients of the $\beta$ function, see
Eq.~(B.13) in~\cite{Andersen:2005bj}.}.
As usual, perturbative QCD calculations
facilitate the determination of the first few orders in
${\cal S}\left(\alpha_s\right)$,
corresponding to the expansion of $B_{\cal S}(u)$ around the origin.
In order to know the pattern of
infrared renormalon singularities of the evolution kernel, one needs
to know $B_{\cal S}(u)$ as an analytic function, and specifically
find its zeros in order to deduce which power ambiguities are missing.
Here the large--$\beta_0$ limit becomes useful: when considering the
evolution kernel a single dressed gluon becomes a natural approximation;
non-Abelian diagrams are expected to modify the residue of the renormalons
by  $C_A/\beta_0$--dependent contributions (${\cal O}(1)$ effects) but
not to alter their positions. Thus, the pattern of zeros of $B_{\cal S}(u)$
can be directly deduced from the large--$\beta_0$ limit.
Table~\ref{Table:Sud} lists the results in some important examples.
For each observable we quote\footnote{Technically, $b_{\cal S}(u)$ is
directly obtained from the Laplace transform of the ``characteristic
function'', which is the leading--order result with a single
off-shell gluon~\cite{Dokshitzer:1995qm,Ball:1995ni}. An example of
such a calculation (for the $C$ parameter in $e^+e^-$ annihilation)
can be found in \cite{Gardi:2003iv}.} the expression for $b_{\cal
S}(u)$ that is related to the corresponding $B_{\cal S}(u)$ by:
\begin{equation}
\label{B_b} B_{\cal S}(u)={\rm e}^{\frac53 u}\,\frac{\sin \pi u}{\pi
u}\, b_{\cal S}(u)\, \times \,V_{\cal S}(u)\,;\qquad \quad
V_{\cal S}(u)= 1+{\cal O}(u/\beta_0),
\end{equation}
where $V_{\cal S}(u)$ has no renormalon singularities nor zeros.
The table also shows where, on the positive real axis,
$\left.B_{\cal S}(u)\right\vert_{{\rm large}\,\beta_0}$ vanishes,
thus eliminating the potential renormalon singularity on the r.h.s.
of Eq.~(\ref{Soft_Borel}). Finally it shows which power ambiguities
do appear. Evidently, the renormalon pattern is different for each
class of distributions.

\begin{table*}[t]
\renewcommand{\arraystretch}{.9}
\caption{Results for Sudakov anomalous dimensions in the
large--$\beta_0$
limit, \eq{B_b}~\cite{DGE_thrust,Gardi:2002bg,Gardi:2001di,CG,GR,Gardi:2003iv,Beneke:1995pq,QD},
and the corresponding power corrections in the evolution kernel,
the r.h.s of \eq{Soft_Borel}.
\label{Table:Sud}\vspace{.2pc}} {\small
\begin{tabular}{cccc}
  \hline
  Observable           & ${\displaystyle b_{\cal S}(u)    }$  & $B_{\cal S}(u)  =0$
for $u>0$ & power corrections
\\
\hline
\\
  \begin{tabular}{l}
  Drell--Yan or Higgs production\\
  near partonic threshold
\end{tabular}
       & 2${\displaystyle \frac{\Gamma^{2}(1-u)}{\Gamma(1-2u)}}$
&$\displaystyle{u=\frac{1}{2},\frac{3}{2},\ldots }$ &
${\displaystyle \left( \frac{\Lambda N}{Q}\right)^k}$,\, $k$ even
\\
\begin{tabular}{l}
Event Shapes ($e^+e^-\to {\rm jets}$)\\
near the two--jet limit \end{tabular}
   \begin{tabular}{c}
  Heavy Jet Mass  \\
  Thrust\\
 $c$ parameter
\end{tabular}
         & \begin{tabular}{c}
         \\
         1\\
         2\\
         2${\displaystyle \frac{\Gamma^{2}(1+u)}{\Gamma(1+2u)} }$
         \end{tabular}
          &$u=1,2,\ldots$&
$\left({\displaystyle \frac{\Lambda N}{Q}}\right)^k$, $k$ odd\\
\\
  \begin{tabular}{c}
Heavy Quark Fragmentation for $x\to 1$  \\
Heavy Quark Distribution
for $x\to 1$ \\
(the hard scale $Q$ is the quark mass $m$)
\end{tabular}
    & ${\displaystyle (1-u)\frac{{\pi u}}{\sin \pi u}      }$
& $u=1$  & $\left({\displaystyle \frac{\Lambda N}{m}}\right)^k$, $k\neq 2$, integer  \\
\\
 \hline
\end{tabular}}
\end{table*}

In order to make practical use of
the DGE formulation for the evolution equation (\ref{Soft_Borel}), one
needs to incorporate the known coefficients of the Sudakov anomalous dimension
 into an analytic function of $u$, namely write an ansatz for
 $V_{\cal S}(u)$ in \eq{B_b}. A concrete example
is $V_{\cal S}(u)=\exp (-y_1 u-y_2 u^2+\cdots)$, where
$y_1=\frac{C_A}{\beta_0}\left(\frac{\pi^2}{12}-\frac13 \right)$ is universal
while $y_i$ for $i\geq2$ are observable--dependent.
These coefficient are determined
order by order in perturbation theory up to the order at which
${\cal S}(\alpha_s)$ has been computed:
from the NNLO result for ${\cal S}(\alpha_s)$ one can determine $y_2$, etc.
 The sensitivity of the resulting Sudakov factor $S_N(Q;\mu_F)$
 to the particular ansatz for $V_{\cal S}(u)$
reflects the remaining theoretical uncertainty due to unknown higher--order
corrections, see e.g. Eq. (3.27)
in Ref.~\cite{Andersen:2005mj} and the discussion that follows.

Examining the Borel integral in \eq{Soft_Borel} one finds that
owing to the
suppression by $\Gamma(-2u)$, the large $u$ region is
irrelevant. For $N\ll Q/\Lambda$ the integral is dominated
by small values of $u$, while for large $N$, of ${\cal O} (Q/\Lambda)$, the intermediate
region, $u\sim \frac12$, and above becomes relevant. In this respect
additional constraints on $B_{\cal S}(u)$ away from the origin
are very useful in extending the
perturbative treatment to larger $N$. Such constraints are provided for example
by the zeros of $B_{\cal S}(u)$ deduced from analytic results in the
large--$\beta_0$ limit. Another example is the
calculation of the renormalon residue at $u=\frac12$ in inclusive decays,
which was first used as a constraint on $B_{\cal S}(u)$
in~Ref.~\cite{Andersen:2005bj}.

For infrared and collinear safe observables such as event--shape
distributions, heavy--quark production cross section
in $e^+e^-$ annihilation, and
heavy--meson decay spectra, one can solve the evolution
equations~(\ref{J_ev_Borel}) and (\ref{Soft_Borel}) and
write a single \emph{factorization--scheme--invariant} Sudakov
factor. For example, in the case of $e^+e^-\longrightarrow b+X$ the
jet mass scale is $Q^2(1-x)$ where $Q$ is the $e^+e^-$
center-of-mass energy, while the soft scale associated with
radiation off the b quark is $m(1-x)$ where $m$ is the b quark mass.
The resulting Sudakov factor is~\cite{CC}:
\begin{eqnarray}
\label{alt_Sud} &&{\rm Sud}(Q,m,N) =\exp\bigg\{
\frac{C_F}{\beta_0}\int_0^{\infty}\frac{du}{u} \, T(u)
\left(\frac{\Lambda^2}{Q^2}\right)^u \times
\bigg[\left(\frac{Q^2}{m^2}\right)^u B_{\cal
S}(u)\left(\frac{\Gamma(-2u)\Gamma(N)}{\Gamma(N-2u)}+\frac{1}{2u}\right)
  \\ \nonumber &&\hspace*{300pt}- B_{\cal
J}(u)\left(\frac{\Gamma(-u)\Gamma(N)}{\Gamma(N-u)}+\frac{1}{u}\right)
\bigg] \bigg\}.
\end{eqnarray}
The corresponding anomalous
dimensions
$B_{\cal S}(u)$ and $B_{\cal
J}(u)$ are now known to
NNLO~\cite{QD,Andersen:2005bj,Korchemsky:1992xv,Melnikov:2004bm,MVV}.
A similar expression, with the~\emph{same}
anomalous dimension functions\footnote{The Sudakov factors of the heavy--quark
distribution and fragmentation functions are {\em the same} to all
orders~\cite{QD}. } holds for inclusive B decay spectra, see~Eq.~(3.43) in
Ref.~\cite{Andersen:2005bj}.

It should be emphasized that \eq{alt_Sud} is an \emph{exact} all--order
formula for the Sudakov factor. In particular, it is \emph{exactly
renormalization--group invariant}.
In contrast with fixed--logarithmic--accuracy
formulations, see e.g.~\cite{Lange:2005yw}, there is no need to introduce
any arbitrary renormalization or factorization scales.
Naturally, uncertainties associated with
unknown higher--order corrections to the anomalous dimensions $B_{\cal
S}(u)$ and $B_{\cal J}(u)$ translate into
uncertainties in the calculation of ${\rm Sud}(Q,m,N)$.
By varying the functional form of $V_{\cal J}(u)$ and $V_{\cal S}(u)$ in
Eqs. (\ref{BJ_full}) and~(\ref{B_b}), respectively, under the available
constraints, one has a handle on the remaining uncertainty of the
\emph{perturbative}
calculation.

Although the DGE formulation and conventional
fixed--logarithmic--accuracy approximations make use of the same
 perturbative calculation of the anomalous dimensions,
the results are typically very
different\footnote{Detailed numerical comparison between DGE with
Principal Value prescription and
fixed--logarithmic--accuracy approximations
was made in several examples, see e.g. Figs. 2
and 3 in Ref.~\cite{DGE_thrust} for the case of the thrust
distribution; Figs. 6 through 8 in Ref.~\cite{CG} in the context
of heavy--quark fragmentation; and Fig 13 in Ref.~\cite{Andersen:2005mj}
in the case of the triple differential rate in
${\bar B}\longrightarrow X_u l\bar{\nu}$ decays.}. These
differences
can be understood in detail by expanding the Borel
function and examining the divergence of the emerging
series~\cite{DGE_thrust,Gardi:2001di,GR}. An important lesson from this
comparison is that the order around which the
renormalon divergence sets in varies significantly with $N$. Thus, there is no
single optimal truncation order that could approximate the Borel sum.
Moreover, when the asymptotic nature of the series sets in early, as occurs
for example for inclusive decay
spectra~\cite{Andersen:2005bj,Andersen:2005mj}, the logarithmic--accuracy
criterion becomes completely useless: at that point the distinction between
logarithmically--enhanced perturbative terms and power
terms\footnote{Consider for example contributions to the exponent in
\eq{alt_Sud} from $u\sim \frac12$.}
is no more clear; separation must then be done without expansion.
We note in passing that fixed--logarithmic--accuracy approximations
have spurious Landau singularities in
moment space which strongly restrict their range of applicability.
Such singularities do not occur upon performing the integral
in~\eq{alt_Sud}.

In order to uniquely \emph{define} the perturbative Sudakov factor, and thus
explicitly \emph{separate} between perturbative and non-perturbative
contributions at the power level,
all renormalon singularities need to be regularized.
The natural regularization is
Principal Value integration. This regularization guaranties that the
Sudakov factor, just like the physical moments,
is a real--valued function of $N$, namely,
\begin{equation}
 {\rm Sud}^{\rm (PV)}(Q,m,N)=\bigg[{\rm Sud}^{\rm (PV)}(Q,m,N^*)\bigg]^*.
\end{equation}
This implies that the corresponding pertubative spectrum,
computed by \eq{inv_Mellin}, is real.

Having defined the perturbative sum we are ready to consider power
corrections. Power corrections on the soft scale $Q/N$ appear
in the exponentiation kernel (\ref{Soft_Borel}) and therefore generate a
multiplicative exponential factor:
\begin{eqnarray}
\label{S_NP}
&&S_N(Q,\mu_F)\,\longrightarrow\, S_N^{\rm PV}(Q,\mu_F)\,\times\,S_N^{\rm NP(PV)}
\left(Q\right);\nonumber \\\nonumber \\
&&{\rm where} \qquad S_N^{\rm NP(PV)}\left(Q\right)=
\exp\left\{-\sum_{k=1,\,\,
k\notin Z_{\cal S}
}^{\infty}\frac{\epsilon_k^{\rm PV}}{k!}
\left(\frac{\Lambda}{Q}\right)^k\,
(N-1)(N-2)\ldots(N-k)
\right\},\\
&&{\rm with}\,\,\qquad  Z_{\cal S}\equiv \left\{z\left\vert
B_{\cal S}\left(u=\frac{z}{2}\right) = 0\right. \right\}\, .\nonumber
\end{eqnarray}
Here non-perturbative corrections are introduced
according to the ambiguity structure. Each
renormalon pole in the evolution kernel (\ref{Soft_Borel}) gives rise
to a power correction with the corresponding $N$--dependent residue.
 Zeros in the anomalous dimension $B_{\cal S}\left(u=\frac{z}{2}\right)$, as
detailed in the third column in Table~\ref{Table:Sud}, imply a missing
renormalon ambiguity, so no corresponding power term is required.
Note that the dimensionless parameters $\epsilon_k^{\rm PV}$ are expected to
be of
${\cal O}(1)$, so high powers in the
sum are inherently suppressed by $1/k!$, a direct consequence of the
$\Gamma(-2u)$ suppression factor in Eq.~(\ref{Soft_Borel}). This means that
truncation of the sum in \eq{S_NP} is likely to be a good approximation up to
high moments and a small number of non-perturbative parameters should suffice.

It is important to emphasize that although $S_N^{\rm NP(PV)}(Q)$ sums up
power corrections, its effect on the spectrum near threshold
is not expected to be small. In general, the leading corrections become large
for~$N\sim Q/\Lambda$.
Upon taking the inverse Mellin transformation
the effect of $S_N^{\rm NP(PV)}(Q)$ can be recast as a convolution with a ``shape function''. Locally,
this convolution generates an effect of ${\cal O}(1)$; the most obvious
example is a shift of the entire distribution by the $k=1$ term
in~\eq{S_NP}. Nevertheless,
the formulation in \eq{S_NP} is significantly more predictive than
the conventional
``shape function'' approach, where the entire ``soft'' function
$S_N(Q,\mu_F)$ is being parametrized. The fundamental difference is that in
\eq{S_NP} the first approximation to $S_N(Q,\mu_F)$ is determined by a
calculation and requires no non-perturbative input. Factorization scheme and
scale dependence are completely avoided.
Moreover, certain features (moments) of this distribution are protected by symmetries
($B_{\cal S}\left(u=\frac{z}{2}\right) = 0$) receiving no non-perturbative
corrections. Finally,
as explained above,
the number of such non-perturbative corrections is limited by the factorial
suppression inherent to high powers in~\eq{S_NP}.

Let us turn now to concrete examples. The fourth column in
Table~\ref{Table:Sud} summarizes the power corrections appearing in the
exponent in each class of inclusive distributions. In the following we
discuss them one by one. We will focus on the comparison between the predicted
power corrections and the ones observed in experiment wherever a
dedicated DGE--based study was done.

\subsection{Drell--Yan Production}

As shown in Table~\ref{Table:Sud}, the case of Drell--Yan production
near partonic threshold provides an interesting example where a
$u=\frac12$ singularity, which could have
appeared~\cite{Korchemsky:1994is,Contopanagos:1993yq} in~(\ref{Soft_Borel}) is not
realized once the detailed dynamics is taken
into account~\cite{Beneke:1995pq}.
Let us explain this conclusion using the DGE
terminology established above~\cite{Gardi:2001di}:
\begin{itemize}
\item{} The vanishing of $\left.B_{\cal
S}^{\rm DY}(u=\frac12)\right\vert_{{\rm large}\,\beta_0}=0$, is
understood to be a \emph{general property} of this anomalous dimension,
namely $B_{\cal S}^{\rm DY}(u=\frac12)=0$ in the full theory.
This means that there is no $u=\frac12$ ambiguity in the
corresponding evolution kernel.
\item{} The perturbative Sudakov exponent
admits \emph{infrared safety at the ${\cal O}(N\Lambda/Q)$
power level}.
Therefore, there is no reason to \emph{expect} any ${\cal O}(N\Lambda/Q)$ non-perturbative
corrections.
\item{} The perturbative Sudakov exponent does have a $u=1$ ambiguity
(see Table~\ref{Table:Sud}),
indicating an ${\cal O}\left((N\Lambda/Q)^2\right)$ power correction.
More generally, Table~\ref{Table:Sud} shows that there are no odd power
ambiguities, while all even powers are present. This renormalon structure
leads to a specific parametrization of power corrections according to
\eq{S_NP}, where the sum is over even values of $k$. In practice one should
truncate this sum, restricting the number of parameters according to the
kinematic range and the required accuracy.
\end{itemize}

As already emphasized, the absence of a particular renormalon ambiguity does not
\emph{imply} the absence of corresponding non-peturbative effects,
and therefore addressing the problem by alternative theoretical means
is useful. In the case of Drell--Yan, an alternative approach based
on joint resummation~\cite{Laenen:2000ij} leads to similar
conclusions~\cite{Laenen:2000hs}: there is no ${\cal O}(N\Lambda/Q)$
correction, nor higher odd powers, while even power
corrections are present.

\subsection{Event--shape distributions}

Event--shape distributions, being infrared and collinear safe jet
observables~\cite{Sterman:1977wj},
provide a unique laboratory for resummation
and power corrections.
Consequently, these observables attracted much theoretical
interest~\cite{Webber:1994cp}--\cite{BM}.
High--quality measurements of event--shape distributions
in $e^+e^-$ annihilation have been performed over a wide kinematic range
by several experiments, providing stringent tests of the theory.

The basic motivation is to understand the effect of
hadronization in a quantitative way. The starting point is a perturbative
calculation of the distribution, where the variable is expressed in terms of
on-shell \emph{quark and gluon momenta}.
The corresponding moments are finite owing to
infrared and collinear safety. The computed spectrum is
then compared with experimental measurements, where the variable is computed
in terms of \emph{hadron momenta}.

One would like to understand the
relation between the perturbative distribution and the measured one through
power corrections. As usual, the main challenge is the description of
the threshold region, the two--jet limit. This was the main motivation
for much of the theoretical work on the subject~\cite{Webber:1994cp}--\cite{BM}, and specifically,
for developing the DGE approach~\cite{DGE_thrust,Gardi:2002bg}.

\begin{figure}[t]
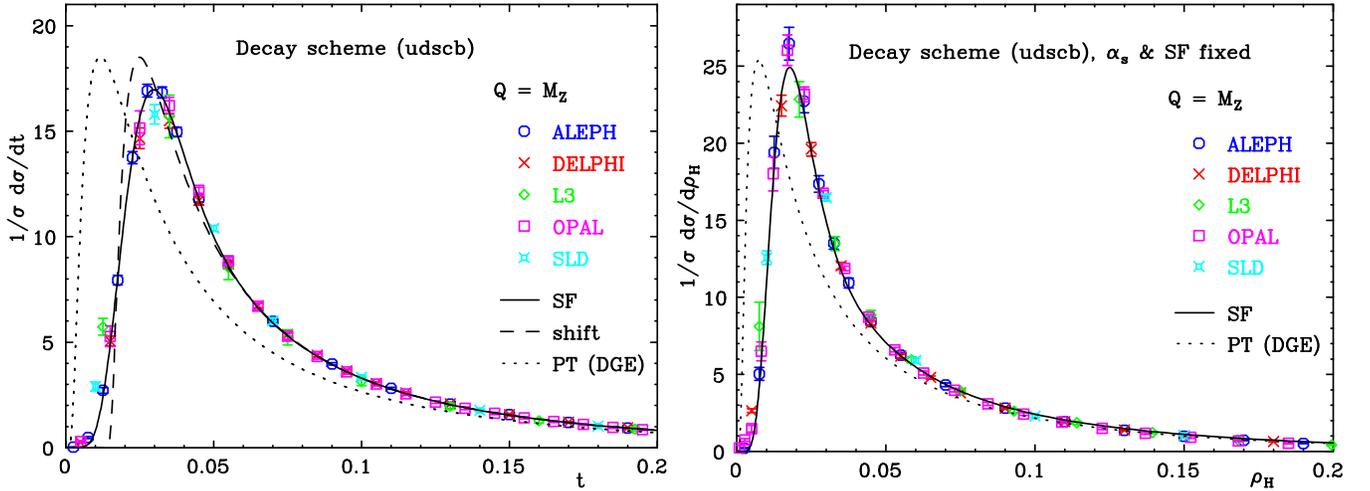

\includegraphics[width=65mm,angle=90]{thrust_kcorr_mz.ps}
\includegraphics[width=65mm,angle=90]{rhoh_kcorr_mz_allfix.ps}
\caption{Event-shape distributions: theory {\it{vs.}} data.
The thrust (left) and the heavy--jet--mass (right) distributions,
computed by DGE~\cite{Gardi:2002bg} for center-of-mass energy of $Q=M_Z$, are compared
with LEP and SLD data. In the thrust case
the three lines show: a Principal Value DGE calculation with no power corrections
(dots); one which is shifted, where the shift parameter is determined is
 a global fit to data (dashed); and one that includes higher power
 corrections, determined in a fit (full line).
 These corrections are found to be consistent with the predication that the
 width does not change due to hadronization. In the heavy--jet--mass case
 no additional
 fits are performed: the dotted line is the result of
 a Principal Value DGE calculation and the full line includes
 hadronization corrections based
 on the fit to the thrust. \label{event_shapes}}
\end{figure}

As shown in Table~\ref{Table:Sud} the single--dressed--gluon calculation
predicts the absence of renormalons at all integer values of $u$; it
therefore predicts only \emph{odd} power corrections
${\cal O}\left((N\Lambda/Q)^k\right)$
in the exponent. The ensuing phenomenology is the following:
\begin{itemize}
\item{} The leading correction
${\cal O}\left(N\Lambda/Q\right)$, generates a global shift of the
distribution and a $\Lambda/Q$ correction to average values of event--shape
variables. This idea~\cite{Webber:1994cp}--\cite{BM} quickly
and led to successful phenomenology: the shift
 can be described in terms of a single, approximately universal
non-perturbative parameter in a large class of
distributions~\cite{Dasgupta:2003iq}.
\item{} When approaching the threshold region, conventional Sudakov resummation
accompanied by a shift of the distribution was proven insufficient. Here
DGE~\cite{DGE_thrust,Gardi:2002bg} made the difference. As shown in
Fig.~\ref{event_shapes} the shape of the (Principal Value) DGE calculation,
before any power corrections are included,
is already very close to that of the data. Specifically,
the \emph{width} of the
distribution is very similar, in accordance with the renormalon prediction
that an ${\cal O}\left((N\Lambda/Q)^2\right)$ correction would not occur.
\end{itemize}

A dedicated DGE--based analysis in Ref.~\cite{Gardi:2002bg} has
demonstrated that the measured hadronization effects are consistent
with the renormalon predictions for power corrections.
Successful fits to experimental data
have been performed over the entire two--jet region and over
a large range of energies. Moreover, as shown in
Fig.~\ref{event_shapes}, the relation
between hadronization corrections to different
shape variables has been tested well into the threshold region.

It is worthwhile noting that the value of $\alpha_s^{\MSbar}(M_Z)$
extracted from fits to
the thrust distribution employing DGE~\cite{Gardi:2002bg}
(or from the average thrust,
employing renormalon resummation~\cite{Gardi:1999dq}) is consistently lower
than the world average value (by about $7\%$). The theoretical
uncertainty (estimated~\cite{Gardi:2002bg} as $\pm 5\%$)
is significantly larger than the experimental one ($\pm 1\%$),
and it is dominated
by uncertainty due to higher--order perturbative corrections. It is therefore
important to revisit the comparison with data once
${\cal O}(\alpha_s^3)$ calculations~\cite{Gehrmann-DeRidder:2006hm} become
available.

An important comment is that in contrast with other distributions considered
here, event--shape distributions are not \emph{completely}
inclusive~\cite{Nason:1995hd,Milan}.
For example, the thrust and the heavy--jet--mass distributions are
defined based on the separation of space into two hemispheres. The
single--dressed--gluon calculation of the Sudakov exponent,
on which the power--correction analysis is based,
assumes that the momentum of final--state particles associated with a
single off-shell gluon is restricted to one hemisphere. This means in particular
that certain correlations between the masses of the two hemispheres are neglected.
The no-correlation assumption is also used when relating~\cite{Gardi:2002bg} the thrust and the
heavy--jet--mass distributions as shown in Fig.~\ref{event_shapes} above.
The successful comparison with data indicates that these
 correlations are indeed small in the two--jet limit.

\subsection{Heavy--quark fragmentation}

Consider next the fragmentation process in which an off-shell
heavy quark of mass $m$,
produced at high energy ${\cal O}(Q)$, fragments into a heavy hadron. This
process is infrared and collinear safe.  Therefore, the fragmentation function
can be computed in perturbation theory, assuming that the final state is an
on-shell heavy quark~\cite{Nason:1996pk,CC,CG,Gardi:2003ar}. This perturbative
fragmentation function differs from the physical one by power corrections
~\cite{Nason:1996pk,CC}.

\begin{figure}[htb]
\includegraphics[width=85mm,angle=90]{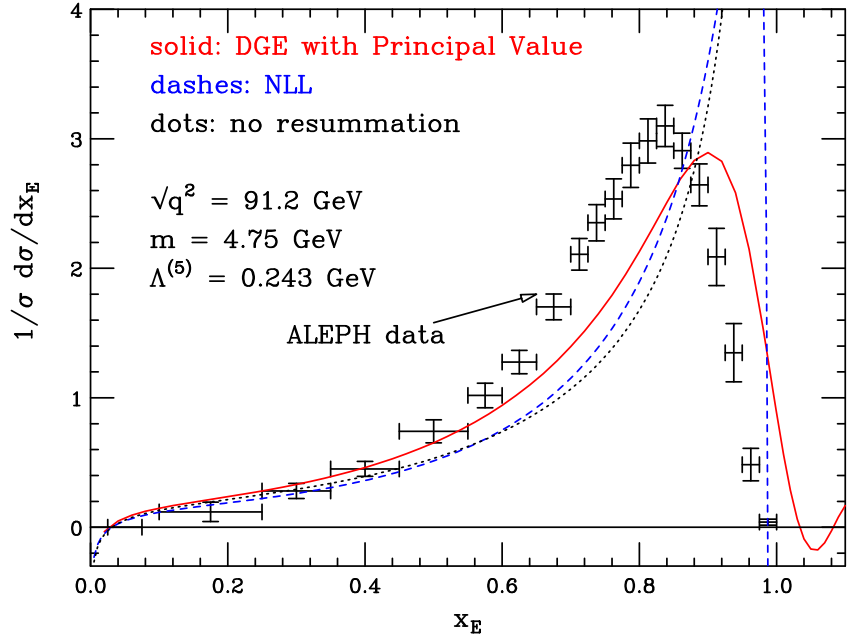}
\caption{Heavy--quark fragmentation: perturbation theory {\it vs.} data.
Comparison between perturbative calculations~\cite{CG} of
the normalized differential distribution in $e^+e^-\longrightarrow b+X$
with respect to $x_E\equiv 2E/Q$, and experimental data from Aleph.
The dotted line is based on a simple NLO calculation with
no soft--gluon resummation: only
evolution logarithms $\ln(m^2/Q^2)$ are resummed; the dashed line
includes in addition soft--gluon resummation to NLL accuracy; the
full line is the corresponding DGE result computed with the
Principal Value prescription (the DGE Sudakov factor
is quoted in \eq{alt_Sud} above).
  \label{heavy_quark_fragmentation}}
\end{figure}

Traditionally, heavy--quark fragmentation effects have been
described in terms of ``fragmentation functions''~\cite{Peterson,Kart},
a given functional form with one or more free parameters.
Upon excluding the difficult threshold region
where the fragmentation function peaks, such fits can indeed
be performed. However, since there is no direct
relation between these models and the QCD definition
of the fragmentation function, the universality of
the extracted parameters is doubtful.
Within the threshold region such fragmentation models simply
fail to bridge the
gap between the resummed perturbative calculation and the data.
This provided a strong incentive to apply the DGE approach to this
problem~\cite{CG,Gardi:2003ar}.

As shown in Table~\ref{Table:Sud}, power--correction analysis of the Sudakov
exponent in the large--$\beta_0$ limit has definite predictions:
\begin{itemize}
\item{} The leading infrared
renormalon is located at $u=\frac12$. Therefore, one expects a global shift of the
distribution in $x_E$ going from the partonic to the hadronic level.
Comparing the data to the Principal value DGE result in
Fig.~\ref{heavy_quark_fragmentation}, one indeed observes such a shift.
The shift parameter $\epsilon_1^{\rm PV}$ in \eq{S_NP}
can of course be determined by fitting the data, as shown in
Fig.~\ref{fig:shiftcurves} below.
\item{} The $u=1$ renormalon is \emph{absent}. Therefore, one expects that the
width of the spectrum would not be modified by non-perturbative fragmentation
effects. Again, this is confirmed by the data~\cite{CG};
see Figs.~\ref{heavy_quark_fragmentation} and \ref{fig:shiftcurves} below.
\item{} Higher renormalon ambiguities at $u=\frac32$ and above are present, and
therefore higher--power corrections are expected. Their effect is
restricted to high moments~\cite{CG}.
\end{itemize}

Fits performed in Ref.~\cite{CG} confirm that \eq{S_NP} above with
a couple of leading non-perturbative power corrections,
yields an excellent description of experimental data to very high moments.
As shown in Fig.~\ref{fig:shiftcurves}, a good description
of data (and a consistent determination of $\alpha_s$)
is achieved already with a single non-perturbative parameter, a plain
shift of the PV-regularized DGE result.
It should be emphasized that such power--correction phenomenology could not
be developed starting with fixed--order calculations or
fixed--logarithmic--accuracy Sudakov resummation: as shown in
Fig.~\ref{heavy_quark_fragmentation} there is a qualitative difference
between these results and the DGE one.

\begin{figure}[t]
\includegraphics[width=165mm]{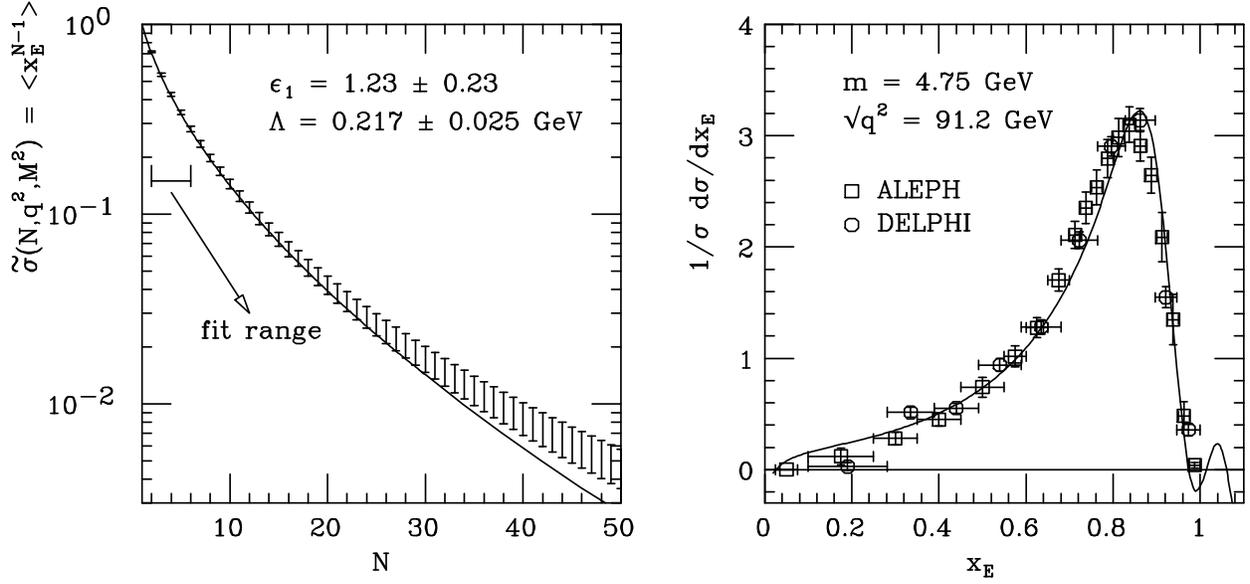}
\caption{\label{fig:shiftcurves} Heavy--quark fragmentation: DGE--based fit
to heavy--meson production data from LEP.
Left: results of a fit
for $\Lambda$ (i.e. $\alpha_s$) and $\epsilon_1^{\rm PV}$ in \eq{S_NP} (setting
$\epsilon_k^{\rm PV} = 0$ for $k\geq 2$)
to Aleph moments $N=2$ to $6$.
Right: the corresponding curve in $x_E$ space, compared to the Aleph
and Delphi data.}
\end{figure}

\subsection{Inclusive decay spectra}

As discussed in Sec.~\ref{BDK}, the application of DGE to inclusive B
decay spectra has been very successful. The most striking fact is that
the on-shell approximation directly provides a good approximation of
the spectrum. In other words, at present accuracy, power corrections on the soft
scale ${\cal O}\left((N\Lambda/Q)^k\right)$, the ones that
distinguish between the quark distribution in a meson and that
in an on-shell quark~\cite{BDK},  are negligible.

To understand this issue, let us return once more to Table~\ref{Table:Sud}:
\begin{itemize}
\item{} As in the case of heavy--quark fragmentation
(as well as in event-shape distributions)
the leading infrared renormalon in the Sudakov exponent
is located at $u=\frac12$. However, a unique property of inclusive decay spectra,
is that this ambiguity cancels out exactly upon computing the resummed spectrum
in physical kinematic
variables~\cite{BDK,Andersen:2005bj,Andersen:2005mj,QD,Gardi:2005mf}.
For example, for the radiative decay the resummed on-shell calculation yields:
\begin{eqnarray}
\label{cancellation}
\left.\frac{1}{\Gamma_{\rm tot}}\frac{d\Gamma}{dE_{\gamma}}
\right\vert_{\rm res}
&=&\frac{2}{m_b}
\int_{c-i\infty}^{c+i\infty}\frac{dN}{2\pi i}
\left(\frac{2E_{\gamma}}{ m_b}
\right)^{-N} \left[H(m_b) \underbrace{J_N(
m_b;\mu_F)\,\times\,S_N(m_b;\mu_F)}_{{\rm Sud}(m_b,N)}\,+\,\,{\cal O}(1/N)\right]
\end{eqnarray}
Here $S_N(m_b;\mu_F)$ is the solution of \eq{Soft_Borel}, the Sudakov factor
describing the momentum distribution in \emph{an on-shell b quark}.
It has a leading ambiguity of the form
$\exp\left\{\pm (N-1)\Lambda/m_b\right\}$.
The mass appearing in the inverse Mellin transformation in (\ref{cancellation})
is the \emph{pole mass},
which also has a leading renormalon
ambiguity~\cite{Bigi:1994em,Beneke:1994sw} at $u=\frac12$.
Since the mass is raised
to the power $N-1$,
it generates an ambiguity of the form $\exp\left\{\mp (N-1)\Lambda/m_b\right\}$
in \eq{cancellation}, which cancels against the ambiguity of
$S_N(m_b;\mu_F)$. This cancellation was confirmed by explicit calculations
of the corresponding renormalon residues in the large--$\beta_0$ limit
(Sec. 2.2.3 in Ref.~\cite{BDK}).\\
\eq{cancellation} refers specifically to the radiative decay, in which
case the meson mass does not enter the calculation.
A similar cancellation occurs in other inclusive distribution, see e.g.
Eq.~(4.4) in Ref.~\cite{Andersen:2005mj} for the charmless semileptonic decay.
There the ambiguity cancels upon expressing the partonic kinematic variables
in terms of hadronic ones, a transformation that involves the mass difference
between the meson and the quark.
\item{} As in the heavy--quark--fragmentation and event--shape--distribution
examples considered above, the $u=1$ renormalon in \eq{Soft_Borel}
is \emph{absent}.
Therefore, one expects that the
width of the spectrum would not be modified by non-perturbative Fermi--motion
effects. Also here, comparison with data confirms the renormalon
prediction:
the photon--energy variance in Fig.~\ref{belle_moments}, computed as
a function of the cut, is well described by the on-shell approximation.
\item{} Finally, there are renormalon ambiguities at $u=\frac32$ and above,
which
imply that non-perturbative power corrections due to Fermi motion
should be included in \eq{S_NP} with $k\geq 3$. These should have an
effect for high moments $N\sim m_b/\Lambda$. Given the factorial suppression
of high powers in \eq{S_NP}, a few non-perturbative parameters would
be sufficient for any practical purpose,
even if very precise data become available.
\end{itemize}

The obvious conclusion is that having sufficient control on the perturbative calculation
and the parametric form of power corrections, the
parametrization of the ``shape function'' and the associated uncertainty
can be completely avoided.

\section{Conclusions\label{Conc}}

The QCD calculation of inclusive cross sections and decay spectra
near kinematic thresholds is challenging because of large perturbative
and non-perturbative corrections.
Conventional Sudakov resummation is limited to the range where
non-perturbative corrections are negligible.
The DGE approach has proven effective in extending the range of
applicability of perturbation theory well into the
threshold region, where non-perturbative corrections are relevant.
DGE makes maximal use of the known all--order structure of the Sudakov
exponent and the inherent infrared safety of the observable.
The main characteristics of this formalism are:
\begin{itemize}
\item{} Factorization and exponentiation in moment space, going beyond the
perturbative (logarithmic) level.
\item{} Resummation of running--coupling effects using
Borel summation, leading to renormalization--group invariance of the Sudakov
exponent.
\item{} Power--like separation between the perturbative result
of the exponent, computed as the Principal Value Borel sum and
non-perturbative power corrections, whose
parametric form is deduced from the renormalon residues.
\end{itemize}
DGE has a high predictive power:
\begin{itemize}
\item{} Important features of the distribution are
captured by resummed perturbation theory.
\item{} The perturbative result
is modified by non-perturbative power corrections of a definite parametric form,
see Eqs. (\ref{J_NP}) and (\ref{S_NP}). The power--correction expansion in
(\ref{S_NP}) converges well even for high moments, and thus a small number of
non-perturbative
parameters is sufficient.
\end{itemize}
The main underlying assumption, whose validity cannot be addressed within
perturbation theory, is that the dominant non-perturbative effects in the
threshold region are indeed the ones identified by renormalon ambiguities.
Here comparison with experimental data is essential.
In all inclusive distributions considered so far, the observed pattern
of power corrections is consistent with the renormalon prediction.
Specifically,
\begin{itemize}
\item{} The potentially leading, $u=\frac 12$ renormalon
ambiguity in the ``soft'' Sudakov exponent $S_N(Q;\mu_F)$ is
of \hbox{${\cal O}\big((N-1)\Lambda/Q\big)$}. This
corresponds to a global shift of the resummed distribution upon going from
the partonic to the hadronic level. This shift has
been firmly established and measured in event--shape
distributions~\cite{DGE_thrust,Gardi:2002bg} (see
also~\cite{Dasgupta:2003iq,Dokshitzer:1997ew,Korchemsky:1995zm}) and in
heavy--quark fragmentation~\cite{CG}. In Drell--Yan or Higgs production the
$u=\frac12$ renormalon is absent due to a specific symmetry property of the
corresponding anomalous dimension, which can
be deduced~\cite{Beneke:1995pq,Gardi:2001di} from \eq{B_b} and Table.~\ref{Table:Sud}
above (see an alternative approach in Ref.~\cite{Laenen:2000ij}).
In inclusive decay
spectra the $u=\frac12$ ambiguity cancels
out within the on-shell approximation,
upon expressing the spectrum in terms of
 hadronic momenta~\cite{BDK,Andersen:2005bj,Andersen:2005mj}.
\item{} The potential $u=1$ renormalon ambiguity does not occur
in several cases,
owing to symmetry properties of the Sudakov anomalous dimensions, which can
be deduced from \eq{B_b} and Table.~\ref{Table:Sud}.
The corresponding non-perturbative
correction would have induced an ${\cal O}(1)$ change in the \emph{width}
of the spectra. Such effects have been excluded by comparison with data in
event--shape distributions (see Ref.~\cite{DGE_thrust,Gardi:2002bg} and
Fig.~\ref{event_shapes} above),
heavy--quark fragmentation (see Ref.~\cite{CG}
and Fig.~\ref{heavy_quark_fragmentation} and \ref{fig:shiftcurves}
above) and inclusive decay
spectra (see Ref.~\cite{Gardi:2005mf} and
the variance plot in Fig.~\ref{belle_moments} above).
\end{itemize}

The recent application of DGE to inclusive decay spectra has led to
a fundamental change in the perception of the predictive power of
resummed perturbation theory. For over a decade, it has been
repeatedly stated that the calculation of the
$\bar{B}\longrightarrow X_s \gamma$ and $\bar{B}\longrightarrow X_u
l \bar{\nu}$ spectra is strictly beyond the limits of perturbation
theory. The description of the spectra in the peak region has been
based on ``shape--function'' phenomenology, where an arbitrary
functional form is introduced whose first few moments are
constrained by data.
Refs.~\cite{BDK,Gardi:2004gj,Andersen:2005bj,Andersen:2005mj,QD,Gardi:2005mf}
have shown that resummed perturbation theory provides a good
approximation to these spectra while genuinely non-perturbative
effects (\eq{S_NP} with $k\geq 3$) due to the ``primordial'' Fermi
motion are not large. This advancement is already reflected in
precision determination of $m_b^{\MSbar}$ and $|V_{\rm ub}|$ using
measurements of the B factories~\cite{HFAG}. Moreover, in this
framework there are high prospects for improving the determination
of these parameters by incorporating higher--order (NNLO)
calculations and by quantifying the relevant non-perturbative
corrections.

\begin{acknowledgments}

I would like to thank Jeppe~Andersen,
Volodya~Braun, Stan Brodsky, Matteo~Cacciari, Stefan~Gottwald,
Georges~Grunberg,
Grisha~Korchemsky,
Lorenzo~Magnea, Johan~Rathsman, Dick~Roberts, Douglas~Ross and \hbox{Sofiane}
 Tafat
for valuable and enjoyable collaboration on various aspects of
renormalons, Sudakov resummation and their applications. I am
grateful to the organizers of the workshops ``FRIF workshop on first
principles non-perturbative QCD of hadron jets,'' January  12 -- 14
2006, Paris; ``Continuous Advances in QCD 2006,'' May 11 -- 14,
2006, Minneapolis; and ``First Workshop on Theory, Phenomenology and
Experiments in heavy flavour physics,'' May 29 -- 31 2006, Capri,
for which this review has been prepared, for the invitation and the
nice hospitality. Finally, I would like to thank Grisha~Korchemsky,
George Sterman and Kolya Uraltsev for illuminating discussions and
Jeppe Andersen and Bryan~Webber for their useful comments on the
manuscript.

\end{acknowledgments}

\end{document}